\documentclass[twocolumn,aps,prd,amsmath,amssymb,preprintnumbers,longbibliography]{revtex4-1}
\usepackage{amsmath} 
\usepackage{amsfonts} 
\usepackage{amssymb}
\usepackage{bbm}
\usepackage{graphics}
\usepackage{graphicx}
\usepackage{titlesec}
\usepackage{mathtools}
\usepackage{environ}
\usepackage{dsfont}

\usepackage[colorlinks=true]{hyperref}
\hypersetup{urlcolor=black,linkcolor=black,citecolor=black}
\usepackage[toc,page,title]{appendix}
\renewcommand{\appendixname}{Appendix}

\usepackage{makecell,tabularx}
\setcellgapes{3pt}

\makeatletter
\renewcommand*\env@matrix[1][c]{\hskip -\arraycolsep
	\let\@ifnextchar\new@ifnextchar
	\array{*\c@MaxMatrixCols #1}}
\makeatother

\newcommand{\be}{\begin{equation}}
\newcommand{\ee}{\end{equation}}
\newcommand{\ba}{\begin{eqnarray}}
\newcommand{\ea}{\end{eqnarray}}
\newcommand{\nn}{\nonumber}

\newcommand{\gl}{\big(}
\newcommand{\gr}{\big)}
\newcommand{\lb}{\left}
\newcommand{\rb}{\right}
\newcommand{\tsigma}{\tilde{\sigma}}

\newcommand{\del}{\partial}

\newcommand{\trho}{\tilde{\rho}}
\newcommand{\veps}{\varepsilon}
\newcommand{\munu}{_{\mu\nu}}

\titleformat{\subsection}[block]{\normalfont\bfseries}{\thesubsection.}{1ex}{}
\titlespacing{\section}{0pt}{20pt}{10pt}[0pt]
\titleformat*{\section}{\large\bfseries}
\renewcommand{\thesubsection}{\arabic{subsection}}

\usepackage{natbib}

\usepackage{braket}

\newcommand{\qq}[1]{``#1''}
\newcommand{\bel}[1]{\be\label{#1}}
\newcommand{\gtil}{\tilde{g}}

\usepackage{todonotes}

\definecolor{refkey}{rgb}{0,0,1}
\definecolor{labelkey}{rgb}{0,1,0}

\renewcommand{\thesection}{\Roman{section}}
\renewcommand{\thesubsection}{\thesection.\arabic{subsection}}

\makeatletter
\renewcommand{\p@subsection}{}
\renewcommand{\p@subsubsection}{}
\makeatother

\usepackage{subfiles}
\usepackage{enumitem}
\usepackage{mathrsfs}

\begin{document}

\title[ ]{Primordial flat frame -- a new view on inflation}

\author{C. Wetterich}
\affiliation{Institut  f\"ur Theoretische Physik\\
	Universit\"at Heidelberg\\
	Philosophenweg 16, D-69120 Heidelberg}

\begin{abstract}
Models of inflationary cosmology admit a choice of the metric for which the geometry of homogeneous isotropic solutions becomes flat Minkowski space in the infinite past. In this primordial flat frame all mass scales vanish in the infinite past and quantum scale symmetry is realized.
The cosmological evolution is dominantly described by the slow increase of a scalar field which sets the scale of all masses. We construct the primordial flat frame for standard models of inflation as Starobinsky inflation or chaotic inflation. In particular, we discuss the evolution of inhomogeneous solutions in the neighborhood of the homogeneous isotropic background solution and their relation to the observable primordial fluctuation spectrum. If the propagators for the graviton and scalar field remain regular, our observed inhomogeneous Universe can be extrapolated back to the infinite past in physical time.  In this case there is no physical big-bang singularity -- the latter reflects only a singular choice of ``field coordinates". Independently of the issue of singularity the primordial flat frame offers a new view on the physical properties of the inflationary universe, which can be characterized as a very slowly evolving almost empty \qq{vacuum} state with approximate scale symmetry.
\end{abstract}

\maketitle

\section{Introduction}\label{Introduction}

Zu Anfang war die Welt oed und leer und waehrte ewig -- in the beginning the Universe was empty and lasted since ever. What sounds like a fairy tale is actually a description of the physical properties of many standard inflationary cosmologies \cite{Starobinsky1980, Guth1981, Mukhanov1981, Linde1982, Albrecht1982, Linde1983, Shafi1983}. In these models the hot big bang is preceded by an inflationary epoch during which almost no particles were present. Particles, radiation and entropy have been created during a heating period after the end of inflation. During inflation, the state of the universe was almost a vacuum. Vacuum is not nothing, however. It is characterized by the expectation values of fields, typically the metric and a scalar field, as well as by the average of the fluctuations of these fields. Propagating fermionic particles are extremely rare during this epoch. All propagating particles behave like photons, traveling with the speed of light. A vacuum with these properties may be called a ``lightlike vacuum".

The physical properties of the early stages of our Universe do not depend on the choice of fields used to describe them. This ``field relativity" \cite{CWUWE,Wetterich2014a} can be used to employ a choice of metric for which the physical properties are particularly apparent. In this ``primordial flat frame" \cite{Wetterich2013,Wetterich2014a} the geometry of homogeneous solutions for the beginning epoch is flat Minkowski space. All masses, both particle masses and the effective dynamical Planck mass, are proportional to a scalar field $\chi$. In the primordial flat frame the evolution of early cosmology is dominantly described by a slow increase of $\chi$.
Somewhat analogously, a beginning with Minkowski space has also been advocated in ``genesis" models based on higher derivative theories \cite{CREM1,CREM2,RUB1,RUB2} or for "stealth solutions" \cite{AMTZ},\cite{FAMO,MOMO}. We do not consider here particular models. We rather
discuss standard inflationary models that can all be formulated with a "beginning" in Minkowski space for a suitable choice of metric.

The primordial flat frame and the standard description of inflationary cosmology in the Einstein frame are related by a Weyl scaling \cite{HW,DIC}, which is a $\chi$-dependent conformal transformation of the metric. Weyl scalings relating different pictures of inflation have been employed since a long time \cite{SW1,SW2}. While the status of field transformations of the classical action is more complex due to Jacobians in the functional integral, the issue becomes very simple on the level of the quantum effective action. Since observables are computed from functional derivatives of the quantum effective action, a change of fields is a simple change of variables in differential equations. All choices of ``field coordinates" are equivalent if the mapping between different choices of fields is invertible. This point of view has been taken in ref.~\cite{CWVN} and should settle a long debate, for a review see ref.~\cite{FAR}. The detailed mapping of many quantities relevant for cosmology has been established \cite{FHU,DAM,FLA,CPS,DESA,CHYA,POVO,NAYOO,DOMNA}. This includes the primordial cosmic fluctuations \cite{Wetterich2015,KPT,BOW,BOOT}, or different time variables \cite{Wetterich2014} -- both particularly relevant for the present work. Observable quantities can be formulated in terms of dimensionless invariants which take the same value in all conformally related frames \cite{CWVN,CPS,Wetterich2015,Wetterich2014,JKSW,JKMR,KAPI}.

In the present paper we construct explicitly the primordial flat frame for many models of inflation. This includes Starobinsky inflation \cite{Starobinsky1980} or chaotic inflation \cite{Linde1983}, as well as large families of neighboring inflationary models. We emphasize the appearance of quantum scale symmetry in the infinite past. For a large family of scaling solutions in quantum gravity the primordial flat frame is computed explicitly. The map between the standard Einstein frame and the primordial flat frame proves very helpful for a focus on quantities that are, in principle, observable. Such ``observable quantities" reflect the physical properties, in distinction to dimensionful quantities as the curvature scalar or the squared Weyl tensor whose possible singularities can be artifacts of a singular choice of field coordinates.

The comparison of different frames for the metric is well suited for a discussion of the character of singularities. Physical singularities arise if the expectation values for physical observables diverge. This often indicates a shortcoming of a model or inappropriate initial conditions, which should be replaced by a more complete model or different initial conditions not leading to a physical singularity. Physical singularities concern observables and do therefore not depend on the choice of fields or the metric frame. If at least one frame exists that is free of singularities, there cannot be a physical singularity. In contrast, field singularities can arise for certain choices of the metric, or more generally “field-coordinates”. They are the analogue of coordinate singularities, but should not be confounded with the latter. They are singularities in field space, rather than in coordinates for a geometric manifold. Field singularities can be removed by a different choice of fields and are therefore not physical singularities. Typically, the transition from a regular choice of fields to a singular choice involves a field transformation that becomes singular at certain points in field space. In the absence of physical singularities the space of fields can be covered by an atlas of charts such that within every chart no singularity occurs. For the absence of physical singularities it is sufficient that for every given observable a choice of fields exists for which the expectation value can be computed and is finite.

A particular emphasis of the present paper concerns inhomogeneous cosmologies. 
Inhomogeneous solutions in the neighborhood of the homogeneous solutions are all attracted towards the homogeneous solution as time progresses. Weak inhomogeneities tend to vanish if the beginning epoch or inflation lasts very long. This can constitute a problem if one tries to extrapolate the observed cosmic inhomogeneities backwards to the infinite past. For certain simple inflationary models the universe would have to start in the infinite past with infinite relative inhomogeneities in order to reach the finite inhomogeneities present at the end of the inflationary epoch.

We formulate the evolution of small fluctuations around a homogeneous isotropic background cosmology in a frame-invariant way \cite{Wetterich2015}. The mode functions appearing in this problem are the same as the ones appearing in the propagator for physical graviton and scalar modes, and therefore in the primordial fluctuation spectrum. This allows us to relate the fate of the relevant inhomogeneous cosmological solutions in the past to the evolution of the primordial cosmic fluctuation power spectrum. A propagator which remains finite for all times including the infinite past implies the presence of regular inhomogeneous solutions 
in the infinite past which can evolve into the observed cosmic inhomogeneities.

In the approximation to the quantum effective action corresponding to standard inflationary models the propagator for the relative graviton fluctuations diverges in the
infinite past. In turn, the inhomogeneous solutions corresponding to the observed primordial power spectrum grow outside the validity of the linear approximation in the infinite past. One may therefore question if the approximation to the effective action remains valid for the beginning of the universe.
In the primordial flat frame a divergent graviton propagator in the infinite past
is related to a vanishing of the effective Planck mass at zero value of the scalar field. For a better approximation one may expect that the
graviton propagator remains finite even for a vanishing scalar field, reflecting the presence of higher derivative terms in the effective action 
for the metric.
These higher derivative terms typically do not affect the homogeneous cosmological solution, but they matter for propagators
and therefore influence the fate of the neighboring inhomogeneous cosmologies. We argue that for a frame-invariant metric the inhomogeneous solutions responsible for the observable primordial fluctuation spectrum remain regular for all times. 
In this case our observed inhomogeneous Universe can be extrapolated backwards to the infinite past without encountering a singularity. 

The absence of a physical singularity for homogeneous cosmology is most easily seen in the primordial flat frame where all geometric quantities remain explicitly finite arbitrarily far in the past. 
Minkowski space is geodesically complete. 
Focusing on physical properties, 
the absence of a physical singularity 
can also be seen in the Einstein frame. Apparent big bang singularities in the Einstein frame are field singularities that originate from a singular choice of field coordinates. Furthermore, the Planck mass is no intrinsic scale. It is introduced only by a specific choice of the metric field.

Neither in the primordial flat frame nor in the Einstein frame there is a singularity for finite $\eta$. The issue of possible singularities concerns the limit $\eta\to-\infty$. For the primordial flat frame this limit is Minkowski space. The geometric singularity structure is identical to Minkowski space -- no geometric singularities occur. There are no curvature singularities, geodesics are complete, and the determinant of the metric remains finite and differs from zero. Possible singularities in the Einstein frame depend on the particular model. For some of the models the curvature scalar or other invariants formed from the curvature tensor diverge. For other models, as for Starobinski inflation, the asymptotics is de Sitter space which is free of curvature singularities. Still, de Sitter space is not geodesically complete and the determinant of the metric vanishes for $\eta\to-\infty$. Part of the researchers associate this with a singularity.

We emphasize in this context that geodesics and their completeness are not frame invariant concepts. They are geometric properties of space-time which are affected by singular Weyl transformations. Massive particles do not move on geodesics if their mass depends on time varying fields. Field-dependent masses produce additional forces acting on particles. We will discuss consequences for physical time in sect.~\ref{Lightlike_vacuum_in_the_Einstein_frame}.

We work within the quantum effective action $\Gamma$ for the metric and a scalar field. All effects of quantum fluctuations are included in the quantum effective action. The quantum field equations are derived as the first functional derivative of $\Gamma$. They are exact. The exact inverse propagator is given by the second functional derivative of $\Gamma$. The primordial fluctuation spectrum can therefore be directly computed for a given form of the quantum effective action \cite{CWQCM}. All observables computed from $\Gamma$ and its functional derivatives are independent of the choice of fields, such that field relativity is realised.

Different inflationary models correspond to different assumptions about the form of the quantum effective action. The present paper makes no attempt to compute the quantum effective action or to take position in favor of one or the other proposed inflationary model. We only discuss the primordial flat frame for different proposed inflationary models and the consequences for the interpretation of their physical properties. We should mention that properties as eternal inflation or self-reproducing universes \cite{VIL2, ALIN1, ALIN2} are, in principle, accessible for a given form of the quantum effective action. This concerns the behavior of solutions of the field equations with sufficiently strong inhomogeneities. The present paper does not address this issue. We remain within a setting of "weak inhomogeneities" that can be treated in a linear approximation. It is conceavable that this applies only to a local region of a universe that is more inhomogeneous on very large scales outside our horizon.

The aim of the present paper is not to propose new models for the inflationary universe. We rather discuss the physical consequences of existing models. The primordial flat frame only changes the field coordinates used to describe these models, not their physical content. It helps to focus on observable quantities, rather than on properties that depend on a given choice of field-coordinates.

In sect.~\ref{Beginning_with flat_spacetime} we discuss a particular effective action for the metric and a scalar field that contains no more than two derivatives and has no problems of stability. The solution of the cosmic field equations derived by variation of this effective action exhibits the geometry of flat Minkowski space in the infinite past. We will later argue that this effective action corresponds precisely to Starobinsky inflation in the Einstein frame. In sect.~\ref{Field_relativity} we construct the general map between the Einstein frame and the primordial flat frame. Sect.~\ref{Inflationary_models_in_the_primordial_flat_frame} discusses the primordial flat frame for chaotic inflation and Starobinsky inflation. In sect.~\ref{Inhomogeneous_Universe} we turn to the inhomogeneous Universe. Sect.~\ref{Quantum_scale_symmetry} discusses quantum scale symmetry and the scaling solutions in the functional renormalization flow of quantum gravity. We construct the primordial flat frame for these scaling solutions. Sect.~\ref{Lightlike_vacuum_in_the_Einstein_frame} discusses the physical properties of the beginning epoch in the Einstein frame. In particular, we show that the effective masses of particles go to zero, and that physical time, as measured by the number of oscillations of wave functions, is infinite as one extrapolates back to the past. Sect.~\ref{Discussion} summarizes our conclusions. Various technical parts are displayed in five appendices. Some of the key ideas of this work are summarized in ref.~\cite{CWGE}, and the present work provides for many of the practical computations underlying ref.~\cite{CWGE}.

\section{Beginning with flat spacetime}\label{Beginning_with flat_spacetime}

In this section we present a model which leads to homogeneous cosmological solutions for which the beginning of the Universe is flat Minkowski space in the infinite past. These solutions do not show any singular behavior. We will later show that this model is equivalent to Starobinsky inflation. The exact equivalence will be given by a conformal field transformation of the metric or Weyl scaling in sects. \ref{Field_relativity}, \ref{Inflationary_models_in_the_primordial_flat_frame}.

We start with an ansatz for the quantum effective action for variable gravity \cite{Wetterich2013}, 
\begin{equation}\label{S1}
\Gamma = \int_x \sqrt{g} \left\{ -\frac{\chi^2}{2} R + \frac{1}{2}(B-6)\del^\mu \chi \del_\mu \chi + \lambda(\chi) \chi^4 \right\},
\end{equation}
where we use a euclidean notation and signature $(-,+,+,+)$, with a factor $i$ arising from $\sqrt{g}$, $g=\text{det}(g_{\mu\nu})$. We first consider a particular model given by
\begin{align}\label{S2}
\lambda &= \lambda_0 (1-W)^2, \qquad W = \frac{3x}{2}\left(1-\frac{5x}{6} \ln \left(\frac{2}{3x}\right)\right), 
\nonumber \\
B &= 6x^2 \left[1- \frac{5x}{3}\left(\ln \left(\frac{2}{3x}\right) - 1\right)\right], 
\end{align}
where
\begin{equation}\label{S2A}
x = \frac{1}{\ln \left(\frac{\mu^2}{\chi^2} + c_t\right)}.
\end{equation}

In this variable gravity model all particle masses are proportional to the scale field $\chi$, such that the ratio of particle mass over
Planck mass remains constant even for a time variation of $\chi$. This includes the confinement scale in strong interactions and the Fermi scale in weak interactions to be proportional to $\chi$. This property constitutes a key difference to Brans-Dicke theories or many
other common variants of scalar tensor theories or induced gravity. 
For the discussion of this paper this difference does not play a role since particle masses matter only for the heating after inflation and subsequent cosmological epochs. It is crucial, however, for a correct description of the matter dominated epochs.

The model \eqref{S1}-\eqref{S2A} is stable despite a negative kinetic term for the scalar field $\chi$. Indeed, for variable gravity, with dynamical Planck mass given by the field $\chi$, the stability condition is $B\geq 0$, $\lambda \geq 0$. We will be interested in small values of $\chi$ where both $x$ and $W$ are small. For $\chi \to 0$ the kinetial $K = B - 6$ approaches the conformal value $K = -6$. Also $\lambda$ approaches a constant $\lambda_0$ and the action becomes invariant under quantum scale transformations. The Planck mass $M$ does not appear as a parameter in the effective action \eqref{S1}, \eqref{S2}. The only scale is $\mu$ which is not related to $M$. The model can be extended to include particle physics, with all particle mass scales given $h_p \chi$, and dimensionless couplings taking values $h_p = m_p/M$, where $m_p$ is the particle mass in the Einstein frame.

Perhaps the particular choice of $B$ and $\lambda$ seems rather special. We will see later that this choice corresponds precisely to Starobinsky inflation. Neighboring models correspond to different models of inflation. The detailed forms of $\lambda$ and $B$ matter for the existence of an exact flat space solution for $t \to -\infty$. 
In other words, the "tuning" or particular choice of the functions $\lambda(\chi)$ and $B(\chi)$ arises from two sources. The first is that we want to describe a particular model in the Einstein frame exactly. The second is that for this model we want to choose a metric frame for which the homogeneous solution becomes Minkowski space in the infinite past. For generic small changes of $\lambda(\chi)$ and $B(\chi)$ neither geometry becomes precisely Minkowski space nor will the model in the Einstein frame be precisely Starobinsky inflation. The physical properties will remain similar, however. The 
qualitative behavior of the functions $\lambda(\chi)$ and $B(\chi)$ is rather simple. The potential is almost a 
$\chi^4$-potential , with a coefficient approaching a constant logarithmically from below as $\chi \to 0$. Also $B$ approaches the conformal
value $B=0$ logarithmically for $\chi \to 0$.

The model \eqref{S2} admits a homogeneous isotropic solution for $t\to -\infty$ for which $H$, $\dot{H}$ and $\chi$ all go to zero. The curvature scalar vanishes in the infinite past. More precisely, we find that in the infinite past $t \to -\infty$ the scalar field $\chi$ approaches zero according to 
\begin{equation}\label{AF1}
\chi(t) = \sqrt{ \frac{3}{\lambda_0}} (t_0 - t)^{-1}.
\end{equation}
The scale factor $a(t)$ of a Robertson-Walker metric for homogeneous isotropic cosmology approaches a constant $\bar{a}$ in this limit
\begin{equation}\label{AF2}
a(t) = \bar{a}\Bigg(1+ \frac{\alpha(t)}{\ln \left(\sqrt{ \frac{\lambda_0}{3}} \mu(t_0 - t)\right)}\Bigg),
\end{equation}
where $\alpha(t)$ is a very slowly varying function. We discuss the field equations derived from the effective action \eqref{S1} and their solution in detail in the appendix \ref{Appendix A}. There we also specify the function $\alpha(t)$.

We can also start in the finite far distant past with an inhomogeneous Universe in the vicinity of the homogeneous isotropic solution \eqref{AF1},\eqref{AF2}. No singularity occurs for increasing time. Such inhomogeneous fluctuations lead to the observed inhomogeneous Universe which can therefore be extrapolated backwards.
The issue of singularities appearing in the infinite past $t \to -\infty$ will be discussed in detail in sect. \ref{Inhomogeneous_Universe}.
Starting arbitrarily far in the past the model is predictive for the properties of fluctuations. There are decaying modes which are predicted to be zero at finite $t$. The primordial scalar and tensor fluctuations of Starobinsky inflation should be non-decaying modes. We discuss in sect. \ref{Inhomogeneous_Universe} the conditions for this property. Starting in the infinite past with finite relative fluctuations of the metric needs the inclusion of higher derivative terms in the quantum effective action. They play no role for the homogeneous solutions. If the conditions for non-decaying models are met, an extrapolation to the infinite past of this model is compatible with all present observations.

For a demonstration that this type of model is generic we may consider a three-parameter family of models that approach scale symmetry logarithmically for small $\chi$. They are given by an expansion
\begin{align}\label{NM1}
\lambda &= \lambda_0 (1 + \tilde{d}_1 x + \tilde{d}_2 x^2 + ...) \nonumber\\
B &= -2\tilde{d}_1 x^2 + \left(\frac{4 \tilde{d}_1}{3} + 2\tilde{d}_1^2 - 4 \tilde{d}_2\right)x^3 +  ...,
\end{align}
with arbitrary coefficients $\lambda_0$, $\tilde{d}_1$, $\tilde{d}_2$. We observe that the expansions of $B$ and $\lambda$ are related. 
This relation reflects the choice of the primordial flat frame for the metric field. In other words, it 
is this relation that ensures that in the infinite past for $t \to -\infty$ Minkowski space is reached, with
\begin{equation}\label{NM2}
\frac{\dot \chi^2}{\chi^4} = \frac{\lambda_0}{3}(1+\tilde{d}_1 x + e_2 x^2 + ...),
\end{equation}
and leading behavior of $\chi(t\to -\infty)$ given by eq.~\eqref{AF1}.
Here the coefficient $e_2$ depends on higher order terms in the expansion \eqref{NM1}. For $t \to -\infty$ the scale factor approaches again a constant according to eq.~\eqref{AF2}, in this case with a constant $\alpha$ given by
\begin{equation}\label{NM3}
\alpha = \frac{3\tilde{d}_2 - \tilde{d}_1-9e_2}{24}.
\end{equation}
We discuss the field equations and solutions of this family of models in the appendix $B$, together with a more general approach to models of variable gravity \eqref{S1} that approach flat space in the infinite past. In the Einstein frame these models are all rather standard inflationary models.

The form of the effective action \eqref{S1} in the ``scaling frame" with dynamical Planck mass given by $\chi$ is preserved by a family of field transformations that combine Weyl scalings of the metric and rescaling of scalar fields \cite{Wetterich2014a,CWVN}. These transformations change $\lambda$ and $B$. They can be used to bring the relation between $\lambda$ and $B$ into a form that admits flat space geometry in the infinite past. 
In particular, for the expansions of $\lambda$ and $B$ in powers of $x$ we can use these transformations to relate the coefficients for 
$\lambda$ and $B$, as in eq.~\eqref{NM1}. We conclude that the existence of such a  ``primordial flat frame" \cite{Wetterich2013, Wetterich2014a} is generic for a very large class of models. The relation between $B$ and $\lambda$ in eq.~\eqref{S2} or eq.~\eqref{NM1} is not a restriction on models. It rather specifies the choice of fields or coordinates in field space for which flat space is reached asymptotically 
in the past.

\section{Field relativity}\label{Field_relativity}

Towards the beginning the ratio between particle mass and momentum goes to zero. All particles become relativistic. This lightlike behavior of all particles towards the beginning of the Universe suggests to use a frame where particle masses are not kept constant, but rather vanish towards the beginning. If the ratio between particles masses and the Planck mass remains constant, the Planck mass also has to vanish towards the beginning. This can be realized by replacing the Planck mass by a scalar field $\chi$, and all particle masses becoming proportional to $\chi$. This is realized by the effective action \eqref{S1} of variable gravity. In such a ``scaling frame" the beginning with vanishing physical particle masses finds a simple description if $\chi \to 0$. We have already seen in eqs.~\eqref{AF1}, \eqref{AF2} that in the scaling frame the big bang singularity is absent for the homogeneous cosmological solution. This demonstrates that in this case the big bang singularity is actually a ``field singularity" due to an inappropriate choice of fields, rather than a physical singularity. In some respect it is analogous to the coordinate singularity at the south pole in Mercator projection coordinates, except that we speak now about ``coordinates in field space".
We will trace the origin of this field singularity by transforming the effective action \eqref{S1} to the Einstein frame, or vice versa.

\subsection*{Weyl scaling}

In the familiar Einstein frame the effective action describing the inflationary epoch involves the metric and a scalar ``inflaton" field $\sigma$, 
\begin{equation}\label{eq:FR1}
	\Gamma = \int_x \sqrt{g_E} \lb\{ - \frac{M^2}{2} R_E + \frac{1}{2} \partial^\mu \sigma \partial_\mu \sigma + V_E(\sigma) \rb\},
\end{equation}
with $V_E$ the effective scalar potential in the Einstein frame. Performing a Weyl transformation we choose a different metric field
\begin{align}\label{eq:FR2}
	g_{E,\mu \nu} = w^2 g_{\mu \nu}, && w^2 = \frac{\chi^2}{M^2},
\end{align}
with $g_{E, \mu \nu}$ and $g_{\mu\nu}$ the metric in the Einstein and scaling frame and $\chi$ a scalar field that will be related to $\sigma$.

Expressed in terms of $g_{\mu \nu}$ the action \eqref{eq:FR1} reads, with $\tilde{\sigma} = \sigma / M$, 
\begin{equation}\label{eq:FR3}
\Gamma = \int_x \sqrt{g} \lb\{ - \frac{1}{2} F(\chi) R + \frac{1}{2} K(\chi) \partial^\mu \chi \partial_\mu \chi + U(\chi)\rb\},
\end{equation}
where
\begin{align}\label{eq:FR4}
	F(\chi) = \chi^2, && U(\chi) &= \lambda(\chi)\chi^4, 
\nonumber\\
	\lambda(\chi) = \frac{V_E(\tilde{\sigma})}{M^4}, && K(\chi) &= \chi^2 \lb( \frac{\partial \tilde{\sigma}}{\partial \chi} \rb)^2 -6.
\end{align}
We assume a monotonic behavior
\begin{equation}\label{eq:FR5}
	\lb( \frac{\partial \tsigma}{\partial \ln\chi} \rb)^2 = B = K+6 > 0.
\end{equation}
During inflation the $\chi$-dependence of $\lambda$ is directly related to the slow roll parameter $\varepsilon$,
\begin{equation}\label{eq:FR6}
	\lb( \frac{\partial \ln\lambda}{\partial \ln\chi}\rb)^2 = B \lb( \frac{\partial \ln V_E}{\partial \tsigma}\rb)^2 = 2B\varepsilon.
\end{equation}
The general solutions of the field equations of ``variable gravity" based on the action \eqref{eq:FR3} are discussed in ref.~\cite{Wetterich2013}.

The Weyl scaling of the metric is accompanied by a rescaling of fermion and additional scalar fields as the Higgs doublet. Every mass in the standard model of particle physics is multiplied in the scaling frame by a factor $\chi/M$. If in the Einstein frame the values of running couplings are defined at a renormalization  scale given by  $M$, the corresponding renormalization scale in the scaling frame is given by the field $\chi$. As a result, all mass scales as the Fermi scale or the confinement scale of strong interactions are proportional to $\chi$. Only mass ratios are observable. They are frame-independent and do not depend on $\chi$. In the scaling frame one finds the quantum scale invariant standard model \cite{Wetterich1988,Shaposhnikov2009,Wetterich2019}. 
For a constant electron mass or proton mass in the Einstein frame, as usually assumed, these masses are proportional to $\chi$ in the primordial flat frame. In particular, the confinement scale and all scattering cross sections or binding energies are proportional to powers of $\chi$, as given by their dimension.
The only violation of quantum scale symmetry is then related to the possible presence of an intrinsic mass scale in the dimensionless functions $\lambda(\chi)$ or $K(\chi)$.

\subsection*{Primordial flat frame}

So far we have not fixed the relation between $\sigma$ and $\chi$. At this stage one has a family of frames according to different possible choices for the relation between $\sigma$ and $\chi$, or the choice of the function $\tsigma(\chi)$. Many models admit a ``primordial flat frame" by a choice of $\tsigma(\chi)$ for which
\begin{align}\label{eq:FR7}
	K<0, && K+6 = \frac{\partial \ln K}{\partial \ln \chi} - \frac{\partial \ln \lambda}{\partial \ln \chi}.
\end{align}
With this choice there are cosmological solutions for which spacetime is flat.

For frames obeying the condition \eqref{eq:FR7} the field equations -- displayed in appendix \ref{Appendix A} in eqs. \eqref{eq:FR9}--\eqref{eq:FR8} -- can be solved for a flat Minkowski geometry, where
\begin{align}\label{eq:FR11}
\dot \chi = \sqrt{ -\frac{2\lambda}{K}} \chi^2, && H=0, && R=0.
\end{align}

For $\chi(t)$ one finds the formal solution
\begin{equation}\label{eq:FR12}
\chi(t) = \chi_0 \lb( 1+ \chi_0 \int_t^{t_0} d t' \sqrt{-\frac{2\lambda}{K}(t')}\rb)^{-1},
\end{equation}
where $\chi_0 = \chi(t_0)$ and $(\lambda / K)(t') = (\lambda/K)(\chi(t'))$. If $2\lambda/K$ reaches a constant $-c^2$ for $\chi \to 0$, one finds for the asymptotic behavior in the past infinity $t\to -\infty$ that $\chi$ vanishes according to
\begin{align}\label{eq:FR13}
\chi(t) \to \frac{1}{c(t_0-t)+\chi_0^{-1}}, && c= \sqrt{ -\frac{2\lambda}{K}}.
\end{align}

The solutions \eqref{eq:FR11}, \eqref{eq:FR12} in the primordial flat frame show no trace of any singularity. All particles become massless in the infinite past as $\chi(t \to -\infty) \to 0$, but massless particles pose no problems in particle physics. The existence of such a frame clearly demonstrates the absence of a physical singularity. For the homogeneous cosmological solution the singularity in the Einstein frame is a field singularity induced by the singularity in the field transformation \eqref{eq:FR2} for $\chi \to 0$.
The metric $g_{\mu\nu}$ amounts to ``regular field coordinates" for the infinite past, whereas the Einstein metric $g_{E,\mu\nu}$ corresponds to ``singular field coordinates". The physical properties are easier to understand by using regular field coordinates. 

We next want to find the relation between $\tsigma$ and $\chi$ for which the ``flat frame condition" \eqref{eq:FR7} holds. This condition can be written as a differential equation for the function $B(\chi)$ or $B(\tsigma)$ namely 
\begin{equation}\label{eq:FR14}
B=2\varepsilon \lb( 1 \pm \frac{1}{\sqrt{2\varepsilon}(6-B)} \frac{\del B}{\del \tsigma} \rb)^2.
\end{equation}
Here the minus sign applies if $V_E$ decreases with $\sigma$ and $\chi$ increases with $\sigma$, while the plus sign accounts for $V_E$ increasing with $\sigma$ and $\chi$ decreasing with $\sigma$. 
Indeed, insertion of eq.~\eqref{eq:FR5} into eq.\eqref{eq:FR7} yields
\begin{align}\label{eq:FR15}
B &= \frac{1}{B-6} \frac{\del B}{\del \ln \chi} - \sqrt{2B\varepsilon} \nonumber \\
&= -\sqrt{B} \left( \sqrt{2\varepsilon} \pm \frac{1}{B-6} \frac{\del B}{\del \tilde{\sigma}} \right),
\end{align}
and taking a square of the last equation implies eq.~\eqref{eq:FR14}.

A primordial flat frame exists whenever for a given $V_E(\tsigma)$ and associated $\varepsilon(\tsigma)$ a solution of eq.~\eqref{eq:FR14} with $0<B(\tsigma)<6$ exists. In particular, for constant $\varepsilon$ one has constant $B= 2\varepsilon \ll 1$, such that $K=B-6$ is indeed negative. For small $\varepsilon$ one finds the iterative solution
\begin{equation}
B = 2\varepsilon \lb( 1 \pm \frac{1}{3-\varepsilon} \sqrt{\frac{\varepsilon}{2}} \frac{\del \ln{\varepsilon}}{\del \tsigma} \rb)^2.
\end{equation}
If the solution of eq.~\eqref{eq:FR14} does not remain within the allowed interval for $B(\tsigma)$ as $\chi$ increases, it is actually sufficient to define $\chi(\tsigma)$ such that the condition \eqref{eq:FR7} holds in the limit $\chi \to 0$. In this case one finds solutions that approach flat space in the infinite past and are again free of singularities.

The strategy for finding a primordial flat frame for a given model of inflation looks first at a (approximate) solution of eq.~\eqref{eq:FR14} for $B(\tsigma)$. Here the particular inflationary model is characterized by the function $\varepsilon(\tilde{\sigma})$. For a given $B(\tilde{\sigma})$ one employs a solution of eq.~\eqref{eq:FR5} for establishing the relation between $\tilde{\sigma}$ and $\chi$. Having done this, one obtains $\lambda(\chi)$ and $K(\chi)$ by inserting $\tsigma(\chi)$. Eq.~\eqref{eq:FR3} with $F = \chi^2$ then defines the effective action in the primordial flat frame. The approximation used for the solution of eqs.~\eqref{eq:FR15} and \eqref{eq:FR5} determines the precise behavior of the cosmological solutions for the field equations derived from the effective action \eqref{eq:FR3}. For an exact solution of the differential equation \eqref{eq:FR5}\eqref{eq:FR15} flat Minkowski space is an exact solution of the field equations, without any time evolution of geometry. If the approximation becomes exact only in the limit $\chi \to 0$, one only can infer that geometry becomes flat space for $t\to -\infty$. For precise enough solutions of eqs.~\eqref{eq:FR15}, \eqref{eq:FR5} in a region of $\chi$ around $\chi = 0$ the geometry will approach a constant scale factor for $t \to -\infty$. For a less accurate approximation one may also have $a(t\to -\infty)\to 0$, with $H(t\to -\infty) = 0$, $\dot H (t \to -\infty) = 0$.

\section{Inflationary models in the primordial flat frame}\label{Inflationary_models_in_the_primordial_flat_frame}

It seems likely that the early stages of all single field inflationary models discussed at present can find a formulation in a primordial flat frame. We see no obvious obstruction to an approximate solution of eq.~\eqref{eq:FR14} for a given $\tsigma$-dependence of the slow roll parameter $\varepsilon(\tsigma)$. A cosmology similar to the one discussed in sect.~\ref{Beginning_with flat_spacetime} is expected provided that inflation has no explicit beginning by a transition from some other type of cosmology. It is instructive to discuss the variable gravity models in the primordial flat frame for specific inflationary models. As examples we take chaotic inflation \cite{Linde1983} and Starobinsky inflation \cite{Starobinsky1980}. In this section we only display the leading order result for the solution $\chi(t)$ for $t\to-\infty$ and compare this with the solution in the Einstein frame. Since conformal time $\eta$ is the same in all frames related by a Weyl scaling, the solutions $\chi(\eta)$ have to be the same in all such frames. A more detailed discussion, including a quantitative computation how geometry deviates from flat space for finite $t$, will be given in various appendices. 

\subsection*{Chaotic inflation}

As a simple example we may consider chaotic inflation \cite{Linde1983}. In the Einstein frame \eqref{eq:FR1} the model is given by
\begin{align}
V_E &= \frac{1}{2}m^2 \sigma^2, \quad \lambda = \frac{1}{2}b \tilde{\sigma}^2, && b = \frac{m^2}{M^2}, \nonumber \\ 
\varepsilon & = \frac{2}{\tsigma^2}, \quad \tsigma = \frac{\sigma}{M}.
\end{align}
For this form of $\veps (\tsigma)$ the primordial flat frame requires $B(\tsigma)$ to obey the differential equation $(\tsigma>0)$
\begin{equation}\label{CI2}
	\frac{\del B}{\del \tsigma} =(6-B)\lb(\sqrt{B} - \frac{2}{\tsigma}\rb).
\end{equation}
For $\tsigma \gg 1$, as appropriate for the inflationary epoch, the solution reads
\begin{equation}\label{CI3}
	B=\frac{4}{\tsigma^2}-\frac{16}{3\tsigma^4}+...
\end{equation}
We choose here the lowest order relation between $\sigma$ and $\chi$ as 
\begin{equation}\label{eq:FR20}
	\chi = \mu \exp \lb(-\frac{\tsigma^2}{4}\rb).
\end{equation}
For large $\tilde{\sigma}$ this obeys eq.~\eqref{CI2}. 

The model that we consider here is therefore given by the effective action \eqref{S1}, with
\begin{equation}\label{eq:FR20A}
B = 2 \ln\left(\frac{\mu^2}{\chi^2}\right), \quad \lambda = b \ln \left( \frac{\mu^2}{\chi^2}\right).
\end{equation}
The Planck mass $M$ is not present in this model. The only intrinsic scale is $\mu$ which is unrelated to the Planck mass. A ``transplanckian" value $\sigma \gg M$ is sometimes considered as problematic. This issue is absent in the primordial flat frame. Large $\tsigma$ corresponds to small $\chi/\mu$. Small field values $\chi \ll \mu$ do not seem to be problematic. The notion of ``transplanckian" fields is a frame dependent issue not directly related to any obvious problem for observables. 

The leading order solution for $t \to -\infty$ is given by eq.~\eqref{eq:FR11}. Thus for $\chi$ close to zero the evolution equation \eqref{eq:FR11} for $\chi$ becomes
\begin{equation}\label{eq:FR21}
	\dot \chi = \sqrt{\frac{2 b}{3}} \chi^2 \sqrt{\ln(\mu/\chi)} = c \chi^2.
\end{equation}
The combination $c=\sqrt{-2\lambda/K}$ approaches for large $\tsigma$ an increasing value
\begin{equation}
c(t) \to \sqrt{ \frac{\lambda}{3}} \approx \sqrt{\frac{ b}{6}}\tsigma(t) \approx \sqrt{ \frac{2 b}{3}\ln \left( \frac{\mu}{\chi(t)}\right)}.
\end{equation}
Correspondingly, $\chi(t\to -\infty)$ reaches zero somewhat faster than $\sim 1/t.$ Replacing the constant $c$ in eq.~\eqref{eq:FR13} by the function $c(t)$ actually becomes a good approximation for $t \to -\infty$.
Relative corrections are proportional to 
\begin{equation}
- \frac{\dot c (t_0-t)}{c} = \frac{1 - \frac{\chi}{\mu}}{2 \ln \left( \frac{\mu}{\chi}\right)}.
\end{equation}

This correction vanishes for $\chi \to 0$. In flat space conformal time is proportional to cosmic time, $\eta = t/a_0$. In the approximation where $a(t)$ can be taken as a constant $a_0$ we find the implicit approximate solution for $\eta \to - \infty$,
\begin{equation}\label{eq:FR23}
\chi^{-1}(\eta) = \sqrt{ \frac{2 b}{3} \ln \frac{\mu}{\chi(\eta)}}\  a_0(\eta_0-\eta) + \chi_0^{-1}.
\end{equation}

We next show that the same solution for $\chi(\eta)$ is found in the Einstein frame if we translate the slow roll solution for $\sigma(t)$ to conformal time and then to $\chi(\eta)$ according to eq.~\eqref{eq:FR20}. As it should be, the behavior of $\chi(\eta)$ does not depend on the choice of frame.

For chaotic inflation in the Einstein frame the field equations read (with $H_E$ the Hubble parameter in the Einstein frame)
\begin{align}
\ddot \sigma + 3H_E\dot\sigma +m^2\sigma = 0 \nonumber \\
3M^2H_E^2 = \frac{m^2}{2} \sigma^2 + \frac{1}{2} \dot \sigma^2.
\end{align}
In the slow roll approximation for large $\tilde{\sigma} = \sigma/M$ they reduce to
\begin{align}
H_E^2 = \frac{m^2}{6}\tsigma^2, \quad 3H_E\dot \tsigma = - m^2 \tsigma,
\end{align}
or
\begin{equation}
\dot \tsigma = - \sqrt{ \frac{2}{3}} m.
\end{equation}
In this approximation the solution reads
\begin{align}
\tsigma = \tsigma_0 + \sqrt{ \frac{2}{3}} m (t_0 -t), \nonumber \\
H_E = \frac{m}{\sqrt{6}}\tsigma_0 + \frac{m^2}{3}(t_0 -t).
\end{align}
Correspondingly, the scale factor obeys \cite{STAR2}, $\bar{a}_0 = a(t_0)$,
\begin{equation}
a(t) = \bar{a}_0 \exp \left\{ - \frac{m}{\sqrt{6}}\tsigma_0(t_0-t) - \frac{m^2}{6}(t_0 -t)^2 \right\}.
\end{equation}

We want to translate this approximate solution to conformal time $\eta$. In the limit of large $\tsigma$ one has $|\dot H_E/H_E^2| \ll 1$ and finds 
\begin{equation}\label{CIE6}
\bar{a}_0( \eta_0 -  \eta) = \frac{1}{H_E(t)} \left[ \exp\left\{ \frac{m}{\sqrt{6}} \tsigma_0(t_0-t) + \frac{m^2}{6} (t_0 -t)^2 \right\} -1 \right],
\end{equation}
or
\begin{equation}
 \eta_0 - \eta = \frac{1}{H_E(t) a(t)} - \frac{1}{H_E(t) \bar{a}_0}.
\end{equation}
As expected, conformal time diverges for $a \to 0$. The physical time elapsed since the big bang singularity at $a = 0$ becomes infinite also in the Einstein frame, see sect.~\ref{Discussion} for a more detailed discussion.
We can write eq.~\eqref{CIE6} in the form
\begin{equation}\label{CIE8}
H_E \bar{a}_0 ( \eta_0 - \eta) +1 = \exp \left\{ \frac{1}{4} (\tsigma^2 - \tsigma_0^2)\right\}.
\end{equation}

We next translate the solution $\tsigma(\eta)$ to $\chi(\eta)$ in order to permit comparison with eq.~\eqref{eq:FR23}.
Using the relation \eqref{eq:FR20} between $\tsigma$ and $\chi$ eq.~\eqref{CIE8} yields
\begin{equation}
\frac{\mu}{\chi} = \exp \left( \frac{\tsigma_0^2}{4} \right) (H_E \bar{a}_0 ( \eta_0 - \eta)+1).
\end{equation}
With
\begin{align}
H_E =m \sqrt{ \frac{2}{3} \ln\left( \frac{\mu}{\chi}\right) }, && \exp \left( \frac{\tsigma_0^2}{4}\right) = \frac{\mu}{\chi_0},
\end{align}
we obtain the implicit equation
\begin{equation}
\frac{1}{\chi} = \sqrt{ \frac{2b}{3} \ln \left( \frac{\mu}{\chi}\right)} \frac{M}{\chi_0} \bar{a}_0( \eta_0- \eta) + \frac{1}{\chi_0}.
\end{equation}
This indeed coincides with eq.~\eqref{eq:FR23} if we identify $a_0 = \bar{a}_0 M/\chi_0$.

The solution \eqref{eq:FR23} reflects only the leading behavior of $\chi(\eta)$ for $\eta \to -\infty$. For corrections to this solution and for a discussion of the associated geometry we present in the appendix C a detailed discussion of the variable gravity model that corresponds to chaotic inflation in the primordial flat frame. 
Exact Minkowski space obtains in the infinite past only if we employ the primordial flat frame relation between $\tsigma$ and $\chi$ beyond the leading order \eqref{eq:FR20}. For the approximation \eqref{eq:FR20} it remains a very good approximation, however.

\subsection*{Starobinsky inflation}

The effective action for Starobinsky inflation,
\begin{equation}
\Gamma = - \int_x \sqrt{g} \left\{ \frac{\mu^2}{2} R + \frac{C}{2} R^2 \right\},
\end{equation}
can be cast into the form \eqref{eq:FR1} by introduction of an explicit scalar field \cite{BWH}, see e.g. \cite{Wetterich2014b}. In this case the potential in the Einstein frame reads
\begin{equation}
	V_E = \frac{M^4}{8C} \left[ 1- \exp\left( -\sqrt{ \frac{2}{3}} \frac{\sigma}{M} \right) \right]^2.
\end{equation}
One infers
\begin{equation}
\lambda = \frac{1}{8C} \left[ 1- \exp \left( - \sqrt{ \frac{2}{3}} \tsigma\right) \right]^2,
\end{equation}
and
\begin{equation}
\varepsilon = \frac{4}{3} \left( \frac{\exp\left(- \sqrt{ \frac{2}{3}} \tsigma\right)}{1 - \exp\left(- \sqrt{ \frac{2}{3}} \tsigma\right)}\right)^2.
\end{equation}

The differential equation \eqref{eq:FR14} for the primordial flat frame can be written in the form
\begin{equation}\label{SI4}
\frac{\del B}{\del\tsigma} = \pm (6-B)(\sqrt{B} - \sqrt{2\varepsilon}),
\end{equation}
where the plus sign applies in our case. For large $\tsigma$ one has
\begin{equation}\label{SI5}
2\varepsilon \approx \frac{8}{3} \exp\left(- \sqrt{ \frac{8}{3}}\tsigma\right),
\end{equation}
and therefore
\begin{equation}
\frac{\del B}{\del \tsigma} \approx 6 \left[ \sqrt{B} - \sqrt{ \frac{8}{3}} \exp \left( - \sqrt{ \frac{2}{3}}\tsigma\right) \right].
\end{equation}
We first consider the leading order solution for large $\tsigma$
\begin{equation}
B = \frac{8}{3} \exp\left(-\sqrt{ \frac{8}{3}}\tsigma\right).
\end{equation}
With
\begin{equation}\label{eq:SI10}
\frac{\del\tsigma}{\del \ln\chi} = - \sqrt{ \frac{8}{3}} \exp\left(-\sqrt{ \frac{2}{3}}\tsigma\right)
\end{equation}
one infers
\begin{equation}\label{eq:SI11}
\exp \left(\sqrt{ \frac{2}{3}}\tsigma\right) = \frac{2}{3} \ln \left( \frac{\mu^2}{\chi^2}\right),
\end{equation}
such that $\varepsilon$ vanishes for $\chi \to 0$ as
\begin{equation}
\varepsilon = \frac{3}{\ln^2 \left( \frac{\mu^2}{\chi^2}\right)}.
\end{equation}

For the asymptotic solution of $\chi(t)$ for $t\to -\infty$ we compute the combination \eqref{eq:FR13},
\begin{equation}\label{eq:SI12}
c \approx \sqrt{ \frac{\lambda}{3}} = \frac{1}{\sqrt{24C}} \left( 1- \frac{3}{4\ln(\mu/\chi)}\right).
\end{equation}
With $c$ approaching a constant the solution approaches eq.~\eqref{eq:FR13} and $\chi$ vanishes in leading order $\chi \sim -1/(ct).$ For Starobinsky inflation in the primordial flat frame the leading behavior for the beginning epoch is flat space with $\chi$ slowly increasing in conformal time $\eta$
\begin{equation}\label{eq:SI13}
\chi = \frac{\sqrt{24C}}{a_0( \eta_0 - \eta)}.
\end{equation} 
Again, the Universe can exist since ever, and all particles become massless in the infinite past for $\eta \to -\infty$

The verification of the asymptotic behavior \eqref{eq:SI13} in the Einstein frame is very simple. For $\tsigma \to \infty$ the geometry for the Einstein frame becomes de Sitter space with constant
\begin{equation}
H_E = \frac{M}{\sqrt{24C}}.
\end{equation}
In the slow roll approximation,
\begin{equation}
\dot \sigma = - \frac{1}{3H_E} \frac{\del V_E}{\del \sigma} = - \frac{M^2}{3\sqrt{C}} \exp \left(-\sqrt{ \frac{2}{3}} \tsigma\right),
\end{equation}
the field $\tsigma$ diverges for $t\to -\infty$ according to
\begin{equation}
\exp\left( \sqrt{ \frac{2}{3}} \tsigma\right) = \frac{M}{3} \sqrt{ \frac{2}{3C}}(\bar{t}-t) = \frac{4}{3} H_E(t_0 -t).
\end{equation}
Switching to conformal time, 
\begin{equation}\label{eq:SI18}
H_E(t_0-t) = \ln \left[ H_E \bar{a}_0(\eta_0-\eta) \right],
\end{equation}
one obtains
\begin{equation}\label{eq:SI19}
\exp \left( \sqrt{ \frac{2}{3}} \tsigma \right) = \frac{4}{3} \ln \left[ H_E \bar{a}_0 (\eta_0 - \eta) \right].
\end{equation}
By use of the relation \eqref{eq:SI11} eq.~\eqref{eq:SI19} becomes 
\begin{equation}
\frac{\mu}{\chi} = H_E \bar{a}_0(\eta_0-\eta).
\end{equation}
With $\bar{a}_0 = a_0 \frac{\mu}{M}$ this is indeed eq.~\eqref{eq:SI13}.

For a connection of Starobinsky inflation to the model \eqref{S1},\eqref{S2} in sect. \ref{Beginning_with flat_spacetime} we have to solve eq.~\eqref{SI4} beyond leading order. An approximate solution is discussed in  the appendix \ref{Appendix D}. It indeed yields the variable gravity model \eqref{S1}\eqref{S2}, for which we discuss the solutions of the field equations in appendix \ref{Appendix A}.

\section{Inhomogeneous Universe}\label{Inhomogeneous_Universe}

Realistic cosmology is not described by a homogeneous and isotropic solution. Inhomogeneities are small at the end of inflation and grow later to form the observed structures as galaxies and clusters. For a discussion of realistic cosmologies we therefore consider for some early epoch relevant for the primordial density fluctuations a geometry with small deviations from a homogeneous and isotropic solution,
\begin{align}\label{IH1}
g_{\mu\nu}(\eta,x) &= a^2(\eta)(\eta_{\mu\nu} + \gamma_{\mu\nu}(\eta,x)), \nonumber \\
\chi(\eta,x) &= \bar{\chi}(\eta) (1 + \delta(\eta,x)).
\end{align} 
Here $\eta$ is conformal time and $x$ are cartesian spatial comoving coordinates. The functions $a(\eta)$ and $\bar{\chi}(\eta)$ are the scale factor and the scalar field according to the homogeneous isotropic ``background solution". The fluctuations $\gamma_{\mu\nu} (\eta,x)$ and $\delta(\eta,x)$ are small and we will linearize in these fluctuations. For a linearized treatment we can deal  separately with fluctuations that belong to different representations of the rotation group since symmetry forbids their mixing. We will work in momentum space with $k$ the three-dimensional comoving momentum. In the linear approximation fluctuation modes with different $k$ cannot mix due to three-dimensional translation symmetry of the background solution. Again, different $k$-modes can be treated separately. A general inhomogeneous Universe \eqref{IH1} is finally obtained as a linear superposition of the different modes.

This is the most general setting for inhomogeneous cosmologies sufficiently close to a homogeneous solution. We do not deal here with strong inhomogeneities, see refs. \cite{EKLS, CLNFFP, CFL} for interesting recent numerical studies. We recall that we do not propose in this paper new inflationary models but rather discuss existing models in different "field coordinates". Solutions of differential equations, including numerical solutions, are the same in all frames. Only the variables are changed. The variables used for the primordial flat frame can shed new light on some of the properties, in particular the connection between the fate of imhomogeneous solutions and the primordial fluctuation spectrum of the associated two point correlation function or propagator.

\subsection*{Graviton fluctuations}

Let us concentrate on the graviton modes which correspond to traceless transversal fluctuations of the metric
\begin{equation}
\gamma_{mn}(\eta,x) = \int_x e^{ikx} \gamma_{mn}(\eta,k),
\end{equation}
with $m,n=1...3$ and
\begin{align}\label{IH3}
k^m  \gamma_{mn}(k) &= k^n \gamma_{mn}(k) = 0, \nonumber \\
\delta^{mn} \gamma_{mn}(k) &=0.
\end{align}
The graviton fluctuations $\gamma_{mn}(\eta,k)$ are relative fluctuations. They are frame invariant quantities, such that their evolution is the same in all metric frames. The associated fluctuations of the metric are given by $g_{mn}=a^2(\eta)\gamma_{mn}$. They depend on the metric frame due to the frame-dependence of the scale factor $a(\eta)$.
For the graviton fluctuations the components $\gamma_{m0}$, $\gamma_{0n}$ and $\gamma_{00}$ vanish and there is no mixing with the scalar fluctuations $\delta$ or other fluctuations of the metric. 
The graviton fluctuations are a particular type of inhomogeneous cosmological solutions in the vicinity of a homogeneous solution.  The behavior of these inhomogeneous cosmologies shows important features that are characteristic for other "weakly inhomogeneous" cosmological solutions.

The field equations for the graviton fluctuations are derived from the effective action \eqref{S1} and read
\begin{equation}\label{IH4}
(\del_\eta^2 + 2 \hat{\mathscr{H}} \del_\eta + k^2) \gamma_{mn}(\eta,k),
\end{equation} 
with
\begin{equation}\label{IH5}
\hat{\mathscr{H}} = \del_\eta \ln(a) + \del_\eta \ln(\bar{\chi}) = a\left(H + \frac{\dot{\bar{\chi}}}{\bar{\chi}}\right).
\end{equation}
The contribution $\sim H$ is the usual Hubble damping, and the term $\sim \dot{\bar{\chi}} /\bar{\chi}$ arises from the $\chi$-dependence of the effective Planck mass in the action \eqref{S1}. (Dots are derivatives with respect to cosmic time $t$.) For the derivation of eq.~\eqref{IH4} one employs the fact that $a(\eta)$ and $\bar{\chi}(\eta)$ are solutions of the field equations. 

For the background solutions in the primordial flat frame we employ, cf. appendix \ref{Appendix A},
\begin{equation}
\frac{\dot{\bar{\chi}}}{\bar{\chi}^2} = \tilde{c}, \quad \frac{H}{\bar{\chi}} = \frac{h_3}{3 \tilde{c}},
\end{equation}
where $\tilde{c}$ is a slowly varying function for $t\to-\infty$, while $h_3$ approaches zero. For the example of Starobinsky inflation one has $\tilde{c} \approx c$ as given by eq.~\eqref{eq:SI12}. In leading order we can neglect in eq. \eqref{IH5} the term involving $H$ and treat $\tilde{c}$ as a constant. This leads to oscillations with a damping proportional to $\bar{\chi}$. To a good approximation the solution for the scalar field reads (cf. appendix \ref{Appendix A}) 
\begin{align}\label{IH8}
\bar{\chi}(\eta) = \frac{1}{a\tilde{c}(\eta_0 - \eta)}, \quad \frac{a\dot{\bar{\chi}}}{\bar{\chi}} = \frac{1}{\eta_0-\eta},
\end{align}
resulting in
\begin{equation}\label{IH9}
\left(\del_\eta^2 + \frac{2}{\eta_0 - \eta} \del_\eta + k^2 \right) \gamma_{mn} = 0.
\end{equation} 
This is the same equation as for the graviton fluctuations in de Sitter space with (almost) constant scalar field where $\hat{\mathscr{H}} = \mathscr{H} = (\eta_0 - \eta)^{-1}$, 
demonstrating once more the equivalence of different frames for the metric.

For a given $k_m$, say $k_m = k \delta_{m3}$, we can group the two modes obeying eq.~\eqref{IH3} into a complex field $\gamma$,
\begin{equation}
\gamma = \gamma_{11} + i \gamma_{12}, \quad \gamma_{22} = - \gamma_{11}, \quad \gamma_{m3} = 0.
\end{equation}
The solution of eq.~\eqref{IH9} reads
\begin{equation}\label{IH11}
\gamma (k) = c^{-}(k) w^{-}_k(\eta) + c^+(k) w_k^+(\eta),
\end{equation}
with mode function
\begin{equation}\label{IH12}
w_k^-(\eta) = (w_k^+(\eta))^* = \frac{\tilde{c}}{\sqrt{2}} k^{- \frac{3}{2}}(-u+i)e^{-iu},
\end{equation}
where
\begin{equation}
u = k(\eta-\eta_0).
\end{equation}
The complex coefficient functions $c^-(k)$, $c^+(k)$ play the role of free integration constants characterizing the different solutions of the linearized equations. Eq.~\eqref{IH11} is the general solution for the Fourier modes $\gamma_{mn}(k)$, with $c_{mn}^\pm (-k) = (c_{mn}^\pm(k))^*$. In the limit considered here $\tilde{c}$ is independent of $u$. The reason for our particular normalization of the mode function $\sim \tilde{c}$ will become apparent below.

The general inhomogeneous solutions \eqref{IH11} are damped oscillations.
In the primordial flat frame this damping is not the geometric Hubble damping due to the expansion of the universe. It is rather induced by the increase of the effective Planck mass due to the increase of the scalar field $\chi$. The 
damping stops for $u^2 \ll 1$. In other words, for $(\eta_0-\eta)^2 \ll 1/k^2$ the amplitude of the fluctuations is ``frozen". This freezing of modes is the analogue of the mode freezing in inflationary cosmology. In the Einstein frame this happens once the wavelength of fluctuations moves outside the horizon. In the primordial flat frame there is no horizon, however, since geometry is Minkowski space. Considerations of causality play no particular role.
The freezing of modes occurs now through the particular dynamics of the increase of the scalar field which constitutes the effective Planck mass.

The damping of the graviton fluctuations demonstrates that the homogeneous solution is a cosmic attractor solution. Small relative inhomogeneities tend to disappear as time progresses. This well known property of inflationary solutions for a fixed Planck mass is reproduced in variable gravity as it should be. If substantial damping lasts for an infinite time, its effect may be so strong that arbitrary finite initial inhomogeneities are completely erased towards the end of inflation. We will discuss this issue in the following.

\subsection*{Graviton propagator}

We have chosen a  normalization of the ``mode function" $w_k^-(\eta)$ such that for Bunch-Davies type initial conditions the equal time graviton propagator reads \cite{Wetterich2015} 
\begin{equation}\label{IH14}
G_{grav} (k,\eta) = 4 |w_k^-(\eta)|^2.
\end{equation}
In our context the graviton propagator is the equal time two-point correlation function. This quantity determines the spectrum of primordial cosmic fluctuations. It should be a well defined quantitiy in any functional integral approach to quantum field theory. There exists an associated operator expression. This is not needed here.
The graviton propagator obtains by inverting the second functional derivative of the effective action \cite{Wetterich2015,CWCF,CWQCM}. Its normalization is therefore constrained. From eq.~\eqref{IH12} one obtains 
\begin{equation}
G_{grav}(k,\eta) = \frac{2\tilde{c}^2}{k^3}(u^2 + 1).
\end{equation}
It is directly linked to the observable tensor power spectrum,
\begin{equation}
\Delta_T^2 (k,\eta) = \frac{k^3}{\pi^2} G_{grav}(k,\eta).
\end{equation}
The tensor power spectrum found in this way will be precisely the same as found in the standard treatment of primordial fluctuations for Starobinsky inflation in the Einstein frame. For $u^2 \to 0$ it reads
\begin{equation}\label{IH16A}
\Delta_T^2 = \frac{2\tilde{c}^2}{\pi^2} \approx \frac{2\lambda}{3\pi^2} \approx \frac{1}{12C\pi^2}.
\end{equation}

The graviton propagator and the evolution of weakly inhomogeneous cosmologies involve the same mode function $w_{k}^{-}(\eta)$. They are therefore closely related. This is not surprising, since the graviton correlation function or propagator can be viewed as a "statistical average" of inhomogeneous cosmologies. This basic property is precisely the content of the functional integral formulation of quantum field theories. The direct relation between the graviton propagator and the evolution of weakly inhomogeneous cosmologies constitutes an important tool for the understanding of the latter. We will exploit this in the following.

\subsection*{Frame invariant evolution equations for \\ fluctuations}

The fluctuation problem can be formulated in a frame invariant way \cite{Wetterich2015}. The fluctuations $\gamma_{mn}(k)$ are the same in all frames. The evolution equation \eqref{IH4} is already written in terms of the frame-invariant quantities
\begin{align}
\label{77}
A = \sqrt{F} a, \quad \hat{\mathscr{H}} =\del_\eta \ln A.
\end{align}
In the scaling frame one has $\sqrt{F} = \chi$, while in the Einstein frame $\sqrt{F}$ is given by $M$.
With $\del_\eta \hat{\mathscr{H}}/\hat{\mathscr{H}}^2 = 1 + \nu$ one finds for $\nu = 0$ the solution
\begin{equation}\label{IH18}
w_k^-(\eta) = \frac{1}{A\sqrt{2k}} \left(1 - \frac{i}{u}\right)e^{-iu}.
\end{equation} 
(For generalizations to $\nu \neq 0$ see ref.~\cite{Wetterich2015}.) The solution \eqref{IH18} is valid in all frames. Both $k$ and $\eta$, and therefore $u$, are invariant under Weyl scalings. For the primordial flat frame has cf.~eq.~\eqref{IH8},
\begin{equation}
\label{79}
A^{-1} = \tilde{c}(\eta_0-\eta) = - \frac{\tilde{c}}{k} u,
\end{equation}
while in the Einstein frame $\chi$ is replaced by the fixed Planck mass $M$ and 
\begin{equation}
\label{80}
A^{-1} = \frac{1}{M a_E} = \frac{H_E}{M} \mathscr{H}_E^{-1} = \frac{H_E}{M}(\eta_0-\eta).
\end{equation}

Using the frame invariant formulation the observable quantities as the fluctuation amplitude or spectral index can be extracted directly. They are obviously the same in the Einstein frame and the primordial flat frame.

The frame invariant expression for the graviton propagator is given by
\begin{equation}
G_{grav}(k,\eta) = \frac{2}{A^2k}\left(1+\frac{1}{u^2}\right).
\end{equation}
In the Einstein frame this yields the familiar expression $(u\to 0,\ A^2u^2 = M^2k^2/H_E^2)$
\begin{equation}
\Delta_T^2 (k) = \frac{2 H_E^2}{\pi^2 M^2},
\end{equation}
where $H_E$ is evaluated at the time of mode decoupling. With $H_E^2 \approx M^2/(24 C)$ this equals the result \eqref{IH16A}.

For the observable modes one has $u^2 \gg 1$ for the early stages before the modes are frozen. (In the Einstein frame this refers to more than 60 e-foldings before the end of inflation.) Following the evolution \eqref{IH11}, \eqref{IH12} backwards for increasing $\eta_0 - \eta$, both the amplitude of the inhomogeneous solutions and the graviton propagator increase. Both quantities remain regular for finite $\eta$, but diverge for $\eta \to -\infty$. 

This behavior constitutes a potential problem. If we start for $\eta\to -\infty$ with finite inhomogeneities, i.e. finite $\gamma_{mn}(\eta\to -\infty,x)$ in eq. \eqref{IH1}, the damping is sufficiently strong to imply for any finite $\eta$ a vanishing inhomogeneity $\gamma_{mn}(\eta,x)=0$. This can be seen directly from the general solution \eqref{IH11} for the differential equation \eqref{IH9}. We may set finite fixed initial conditions $\gamma(k)$ at some time $\eta_{\text{in}}$. This fixes the constants $c^{\pm}(k)$ as a function of $\eta_{\text{in}}$. Keeping the same $\gamma(k)$ for $\eta_{\text{in}}$ assuming more and more negative values the constants $c^{\pm}(k)$ scale with the inverse powers of the mode functions $w_{k}^{\pm}(\eta_{\text{in}})$. For $\eta_{\text{in}}\to -\infty$ the coefficients $c^{\pm}(k;\eta_{\text{in}}\to -\infty)$ vanish. This implies for any finite $\eta$, with finite non-zero $w_{k}^{\pm}(\eta)$, a vanishing inhomogeneity $\gamma(k,\eta)=0$. One finds precisely this behavior for a numerical solution of eq. \eqref{IH9} if one fixes initial conditions at $\eta_{\text{in}}$, "measures" $\gamma_{mn}(\eta)$ at fixed $\eta$, and moves $\eta_{\text{in}}\to -\infty$.

The strong damping of inhomogeneities has a direct consequence for the possibility to extrapolate inhomogeneous cosmologies backwards in time. For obtaining at finite $\eta$ any non-zero inhomogeneity $\gamma(\eta)$, one has to start with an infinite inhomogeneity $\gamma(\eta_{\text{in}})\to\infty$. In turn, this implies that all cosmological solutions with finite inhomogeneities in the traceless transverse tensor sector of the metric will lead to a divergent inhomogeneity when extrapolated backwards to $\eta\to -\infty$. Such solutions become singular in this limit.

\subsection*{Absolute and relative inhomogeneities}
\label{subsec: Absolute and relative inhomogeneities}

The divergence of the graviton propagator concerns the relative fluctuations of the metric. Only those are frame invariant. The fluctuations of the metric itself depend on the frame. They obtain by multiplication of $\gamma_{mn}$ by a factor $a^{2}$. In the primordial flat frame this does not change much since $a(\eta\to -\infty)$ is constant.
In the Einstein frame, however, $a(\eta\to -\infty)$ vanishes. The function $a^{2}(\eta)w^{\pm}_{k}(\eta) \sim \eta^{-2}w^{\pm}_{k}$ is proportional to $a(\eta)$ and therefore vanishes for $\eta\to -\infty$. In this frame the metric fluctuations vanish instead of diverging. The divergence of $\gamma_{mn}$ arises here from the fact that the homogeneous isotropic metric vanishes even faster.

This points to the possibility of a beginning for which the universe is inhomogeneous, but the inhomogeneities and associated two-point functions remain finite. If in this beginning universe the homogeneous expectation values vanish faster than the inhomogeneities for $\eta\to-\infty$, the relative inhomogeneities necessarily diverge. In this case the divergence of the relative inhomogeneities is not associated to a divergent metric, but rather due to the vanishing of the homogeneous expectation value or average of the metric.

For an investigation of this issue we need a frame invariant formulation for the (absolute) inhomogeneous metric. A frame invariant metric can be defined by multiplication with $F(\chi)$,
\bel{AR1}
\gtil_{\mu\nu}=Fg_{\mu\nu}\ ,
\ee
where $F$ is defined by the coefficient of the curvature scalar in the effective action~\eqref{eq:FR3}. The transformation of $F$ under Weyl scalings compensates the one for $g\munu$. With $Fa^2=A^2$ according to eq.~\eqref{77} our solution for the homogeneous frame invariant metric $\gtil\munu$ reads
\bel{AR2}
\gtil\munu=A^2(\eta)\eta\munu\ .
\ee
On the other hand, one obtains for the inhomogeneous metrics from eq.~\eqref{IH18}
\bel{AR3}
\gtil\munu(k)\sim A^2(\eta)w^-_k(\eta)\sim\frac{A(\eta)}{\sqrt{2k}}\gl1-\frac{i}{u}\gr e^{-iu}\ .
\ee
The frame invariant scale factor $A(\eta)$ vanishes for $\eta\to-\infty$ according to eq.~\eqref{79} or~\eqref{80}. We conclude that the frame invariant absolute inhomogeneous metrics $\gtil\munu(k)$ actually vanish for $\eta\to-\infty$, rather than diverging. Only the homogeneous average~\eqref{AR2} vanishes even faster, resulting in divergent relative fluctuations.

For the effective action that is assumed for standard inflationary models as Starobinsky inflation or chaotic inflation the likely beginning of the universe is an inhomogeneous universe with a vanishing homogeneous average value of the metric. Vanishing expectation values are quite common in quantum field theories. Only the geometric interpretation may become problematic. A typical metric of this inhomogeneous universe may be constant for $\eta\to-\infty$, or vanish. It is not sure if the average over the inhomogeneous fluctuations vanishes in the sense that the two-point correlation function vanishes as suggested by eq.~\eqref{AR3}. With a frame-invariant vierbein that remains constant for the linear approximation for $\eta\to-\infty$ non-zero inhomogeneous geometries seem rather plausible. In any case, there is no indication that the frame invariant metric becomes singular for $\eta\to-\infty$.

For the inhomogeneous universe at the beginning stage the primordial flat frame may not be an optimal choice. While the frame invariant scale factor $A(\eta)$ vanishes for $\eta\to-\infty$, the scale factor $a(\eta)$ in the primordial flat frame is kept fixed. As a result, the frame-invariant divergence of the relative inhomogeneities translates into a divergent metric. It is well conceivable that the primordial flat frame helps to understand the physics of the homogeneous universe, while the inhomogeneous universe is better understood in a frame for which $a(\eta)$ vanishes for $\eta\to-\infty$.

\subsection*{Validity of linear approximation}
\label{subsec: Validity of linear approximation}

For the inhomogeneous solutions one may question the validity of the linear approximation since $|\gamma|^2$ exceeds one before growing even further. The linear approximation assumes $|\gamma|\ll 1$, such that the inhomogeneous universe in the beginning stage is not within the validity of linear perturbation theory. Despite this shortcoming, the argument for diverging relative fluctuations seems to hold beyond the linear expansion. Indeed, the correlation function $G_{grav}$ shows the same increase towards a singularity for $\eta\to-\infty$. No linear approximation is used for the inversion of the second functional derivative of the effective action such that the graviton propagator is non-pertubative in this sense. It only employs a given assumed form of the effective action. There is, indeed, a parallel between the correlation function and the behavior of inhomogeneous solutions. A given correlation function is associated with the possible existence of inhomogeneous solutions for which $|\gamma^2|$ is at least of the same order as $G_{grav}$.

One may also question if for leading inhomogeneous metrics the evolution equations for the homogeneous average value remain valid. This concerns the possible \qq{backreaction} of the inhomogeneities on the homogeneous evolution equation. The backreaction can be encoded in an effective energy momentum tensor in the gravitational field equations, and a similar source term for the scalar field equation. We have not investigated if in the presence of backreaction the behavior $\gtil\munu\sim A^2\eta\munu$ in eq.~\eqref{AR2} remains valid. In any case, it seems unlikely that the backreaction of non-singular inhomogeneous fields leads to a singularity in the average of the fields.

\subsection*{Finite relative inhomogeneities?}

We have directly connected the divergence of the relative inhomogeneities to the behavior of the propagator for relative graviton fluctuations $G_{grav}$. We may therefore question if this feature is expected to hold in a more general gravitational effective action beyond the second order in the derivative expansion.

In a more complete description of the quantum effective action one expects higher order derivative terms as
\begin{equation}\label{IH23}
\Gamma_D = \frac{1}{2} \int_x \sqrt{g} C_{\mu\nu\rho\sigma} D C^{\mu\nu\rho\sigma},
\end{equation}
where $C_{\mu\nu\rho\sigma}$ is the Weyl tensor and $D$ may be a function of $\chi$ and the covariant Laplacian. For constant $D$ the term \eqref{IH23} would contribute to the inverse graviton propagator a piece $\sim D q^4$ with  $q^2$ the invariant squared four-momentum in Minkowski space, and a suitable covariant generalization for other geometries. Even though no constant $D$ is expected -- this would induce a ghost-like instability --any smooth function $D$ of the covariant Laplacian would give a modification of the inverse propagator of a similar type, with $D$ replaced by $D(q^2)$. As $\chi \to 0$ the contribution to the inverse propagator $\sim D q^4$ dominates over the contribution from the action \eqref{S1} which is $\sim \chi^2 q^2$. A graviton propagator $G_{grav} \sim D^{-1} q^{-4}$ for $\chi \to 0$ does no longer diverge in this limit for any non-zero $k$.

For a given $k$ the space-inhomogeneity of the graviton solutions contributes to $q^2$ a piece $k^2/a^2$. The term involving $D$ will dominate the graviton propagator provided
\begin{equation}
\frac{D k^2}{a^2} \geq \bar{\chi}^2.
\end{equation} 
With eq.~\eqref{IH8} this results in
\begin{align}\label{IH25}
Dk^2 \gtrsim \frac{1}{\tilde{c}^2 (\eta_0-\eta)^2}, \quad u^2 \gtrsim \frac{1}{D\tilde{c}^2}.
\end{align}
In the range \eqref{IH25}the increase of $G_{grav}$ with increasing $\eta_0-\eta$ will be stopped and $G_{grav}$ is expected to reach a finite value in the infinite past for $\eta \to-\infty$. The inequality \eqref{IH25} can also be seen in the Einstein frame, realizing that the term $\Gamma_D$ is frame invariant for constant $D$. For $D k^2/a_E^2 \gtrsim M^2$ the second functional derivative $\Gamma_D^{(2)}$ dominates the inverse graviton propagator. This again yields the second condition \eqref{IH25}. 

We conclude that for a given $k$ the effective action \eqref{S1}(or similar for other models) can only be used for values of $\eta$ in the range
\begin{equation}\label{85A}
\eta^2 \leq \frac{1}{D \tilde{c}^2 k^2}\,.
\end{equation}
While this range extends to $\eta \to -\infty$ for the homogeneous solution $k \to 0$, it is finite for the inhomogeneous solutions corresponding
to finite $k$. For the observable modes the range \eqref{85A} typically extends far larger than the range needed for the predictions of the 
primordial fluctuation spectrum. (In the Einstein frame the range \eqref{85A} extends much further than 60 e-foldings before the end of inflation.) This statement depends, however, on the value of $D$ and breaks down for very large $D$. Going back in time beyond the limit 
\eqref{85A} one needs to take into account the effect of the term \eqref{IH23}.

The modification of the field equations due to the additional term $\Gamma_D$ will also be reflected in the behavior of the inhomogeneous solutions for $\eta \to -\infty$. In the linear approximation the relations \eqref{IH11},\eqref{IH14} do not depend on the precise form of the effective action. Only the precise shape of the mode function $w_k^-(\eta)$ will be modified since it obeys a modified differential equation. Eq.~\eqref{IH12} will change its behavior in the range $u^2 \gtrsim (D\tilde{c}^2)^{-1}$. The mode function reaches a finite value for $\eta \to -\infty$ whenever $G_{grav}$ remains finite. The backward extrapolation of the observable graviton inhomogeneities then remains  finite in the infinite past. 

This discussion leads to an important conclusion: Whenever the graviton propagator remains finite, also the inhomogeneous graviton solutions for the relative fluctuations remain finite. If a finite graviton propagator can be extrapolated backwards to the infinite past, the same holds for corresponding inhomogeneous solutions of the field equations. 

Strictly speaking, a proof of these statements only holds for solutions that remain within the validity of a linear approximation. Only in this case the evolution of the fluctuations is guaranteed to be given by the same modification of the mode function as the one appearing in the evolution of the propagator. It seems likely, however, that the linear approximation is not essential for the argument. A given fluctuation solution could move outside the domain of validity of the linear approximation as it is extrapolated backwards, and perhaps even diverge. It seems rather unlikely, however, that all generic solutions can diverge. This would require that divergent fluctuations produce a finite correlation function that equals the propagator, $G_{grav}(\eta,x,y) \sim \braket{ \gamma(\eta,x)\gamma(\eta,y)}_c$. 
Except for unnatural cancellations we conclude that there must exist sufficiently many fluctuation solutions -- those that dominate the correlation function -- which remain finite if the propagator remains finite. This argument does not only concern the relative graviton fluctuations. It applies equally to the full metric propagator or the propagator for the frame invariant metric. A finite metric propagator for $\eta \to -\infty$, as expected for any realistic extended form of the effective action, implies that the graviton solutions that are responsible for the primordial graviton spectrum remain finite as well.

\subsection*{Primordial power spectrum}

Our discussion of the graviton fluctuations can be extended to the fluctuations which produce the primordial fluctuations in the scalar sector. Again, a finite propagator in the scalar sector for $\eta \to -\infty$ implies that the scalar fluctuations responsible for the primordial power spectrum remain finite. 
The scalar fluctuations are less problematic than the graviton fluctuations. The scalar kinetic term does not vanish for $\chi\to 0$, such that the scalar propagator in Minkowski space remains finite in this limit. Invariants of a similar type as eq. \eqref{IH23}, with Weyl tensor replaced by the curvature scalar $R$, contribute to the scalar propagator but not to the graviton propagator. Invariants for which the Weyl tensor is replaced by the Ricci tensor $R_{\mu\nu}$ are also expected. These additional invariants do not change the qualitative picture if no ghost of tachyonic instabilities are introduced. We refer to refs. \cite{Wetterich2014a},\cite{Wetterich2019} for a more extended discussion.

The observable primordial power spectrum is the same in the Einstein frame and the primordial flat frame. Our overall conclusion is that the fluctuation solutions responsible for the observable primordial fluctuation spectrum can be extrapolated backwards to the infinite past without encountering any singularity. This holds provided that the effective action for gravity remains regular in the sense that if does not produce singular propagators, except for the expected singularities for particle poles. The observed inhomogeneous Universe remains then regular for all times from the infinite past to the infinite future. The big bang singularity is an artifact of a choice of fields in the Einstein frame that may become singular for $\chi \to 0$ even for the homogeneous cosmologies.

\subsection*{Squared Weyl tensor}

In the Einstein frame it has often been found that for inhomogeneous or anisotropic cosmologies the squared Weyl tensor diverges as $a_E$ goes to zero. From this observation a physical singularity was inferred, arguing with the invariance of the squared Weyl tensor under conformal transformations. This argument is not accurate, however.

The quantity that is invariant under conformal transformations (Weyl scalings) is 
\begin{equation}
W = \sqrt{g} C_{\mu\nu\rho\sigma} C^{\mu\nu\rho\sigma} = \sqrt{g} C^2.
\end{equation}
The factor $\sqrt{g}$, which is often omitted in this type of arguments, plays an important role. We notice that the quantity transforming as a total derivative under general coordinate transformations is also $W$, rather than $C^2$.

Assume now that for some inhomogeneous solution in the primordial flat frame one finds a finite $C^2$ for $\eta \to -\infty$, and that $C^2$ remains finite for increasing $\eta$. In the primordial flat frame also $W$ remains finite, since $\sqrt{g} = a^4$ reaches a constant $\bar{a}^4$ for $\eta\to-\infty$. The frame invariance of $W$ implies that $W$ remains finite in all frames related by Weyl scalings, including the Einstein frame. With $\sqrt{g} = a_E^4$ for the Einstein frame one has
\begin{equation}\label{86}
C_E^2 = \frac{W}{a_E^4}.
\end{equation} 
The squared Weyl tensor $C_E^2$ will diverge for $a_E \to 0$ for all geometries for which $W_\infty = W(\eta \to-\infty)$ differs from zero. Thus a divergence of the squared Weyl tensor is indeed expected for many solutions of the field equations that are not conformally flat. This divergence does not indicate any physical singularity. We have already seen the existence of other choices of fields -- the primordial flat frame -- for which $C^2$ remains regular provided $W_\infty$ remains finite. Again, the singularity of inhomogeneous solutions in the Einstein frame is an artifact due to a singular choice of field coordinates. 

\subsection*{Arrow of time}

Not every arbitrary inhomogeneous Universe can be extrapolated backwards to the infinite past without encountering a singularity. Generic 
inhomogeneous and anisotropic solutions diverge when extrapolated backwards to $\eta \to -\infty$ \cite{KAM2,MRST}. This may be the case even
if the effective action is modified by terms as $\Gamma_D$ in eq.~\eqref{IH23}. The issue is related to the presence of decreasing fluctuation modes. We denote the amplitude of such a decreasing mode by $\varphi(\eta,k)$, that we take real and positive for convenience. For a decreasing mode any interval of values at time $\eta_1$, $\varphi(\eta_1,k) < \bar{\varphi}(\eta_1,k)$, is mapped to a smaller interval at $\eta_2 > \eta_1$, $\varphi(\eta_2,k) < \bar{\varphi}(\eta_2,k)$, $\bar{\varphi}(\eta_2,k) < \bar{\varphi}(\eta_1,k)$. Assume now that for $\eta_1 \to - \infty$ arbitrary values of $\varphi(\eta_1,k)$ (e.g. $\bar{\varphi}(\eta_1,k)\to \infty$) are mapped to a finite interval $\bar{\varphi}(\eta_2,k)$ at $\eta_2$. Starting at $\eta_2$ and following the evolution backwards to $\eta < \eta_2$, only the amplitudes in the interval $\varphi(\eta_2,k) < \bar{\varphi}(\eta_2,k)$ can be followed consistently to $\eta \to -\infty$. In contrast, for all values outside the allowed interval,  $\varphi(\eta_2,k) > \bar{\varphi}(\eta_2,k)$, the backwards solution has to diverge for some finite $\eta_s$. The solution becomes singular, and this singularity cannot be removed by field redefinitions.

This type of singularity does not indicate a singular cosmology. It rather indicates a \textit{prediction} of a certain cosmology, namely $\varphi(\eta_2,k) < \bar{\varphi}(\eta_2,k)$. Universes for which at $\eta_2$ the prediction is violated are not allowed. If one tries, nevertheless, to extrapolate the forbidden cosmologies backwards, the singularity at $\eta_s$ reminds us the inconsistency of the ``forbidden Universes". The situation is analogous to a damped pendulum. If arbitrary initial conditions lead to a maximal amplitude of 1cm after an hour, the attempt to follow an oscillation with amplitude 5cm backwards will lead to a solution that becomes singular in less than an hour backwards.

We emphasize that this type of ``singularity" can only occur for decreasing modes. Decreasing modes often vanish at finite $\eta_2$ if initial conditions with finite $\bar{\varphi}(\eta_1,k)$ are set for $\eta_1 \to -\infty$. In contrast, modes with increasing, constant or almost constant amplitude can be extrapolated backwards to the infinite past. The ``almost constant modes" can decrease for a certain time interval, but they do not vanish at finite $\eta_2$ for $\eta_1 \to -\infty$ and finite $\bar{\varphi}(\eta_1,k)$. They correspond precisely to the ``observable modes" in the primordial fluctuation spectrum. The gauge invariant scalar modes typically contain an almost constant mode as well as decreasing modes. Examples for decreasing scalar modes are discussed in ref.~\cite{Wetterich2013}. They correspond to homogeneous isotropic cosmologies attracted 
towards our "eternal" solutions as time increases. Typically, such deviations from the attractor solution will diverge after a finite time
$\eta_s$ backwards, if they differ from zero at finite time $\eta_2$. In contrast, the  almost constant mode is responsible for the scalar part in the primordial fluctuation spectrum. For the fluctuations of the scale invariant metric the graviton may be an increasing mode if eq.~\eqref{AR3} remains valid, or perhaps turn to a constant mode beyond the approximation used here. There are other fluctuations in the metric that are decreasing modes.

Setting all decreasing modes to zero at some time $\eta_2$ during inflation, the Universe remains inhomogeneous. All types of inhomogeneities that can be accounted for by the almost constant modes are allowed. If the corresponding propagators remains finite, this type of inhomogeneous Universe can be extrapolated to the infinite past without encountering a singularity, similar to the special case of the graviton fluctuations discussed above. The Universe with vanishing decreasing modes is precisely the observed inhomogeneous Universe. This inhomogeneous Universe can then have lasted since the infinite past.

The presence of decreasing modes or damped fluctuations constitutes an arrow of time \cite{Wetterich2013}. While field equations are time reversal invariant, a given homogeneous isotropic solution for the average metric is not. A given non-static cosmology can be viewed as spontaneous breaking of time reversal symmetry. Fluctuations around a given homogeneous isotropic ``background solution" define an arrow of time. The positive time direction is the one for which the decreasing modes get smaller. The presence of an arrow of time is a general property of fluctuations around a time dependent cosmological solution. It does not need concepts as increasing entropy, which does not play an important role in the lightlike vacuum at the beginning of the Universe. Later on, after the end of inflation, entropy increases in the positive time direction that is defined by the behavior of fluctuations.

We conclude that a requirement that arbitrary inhomogeneous solutions can be extrapolated backwards to infinite time for a singularity free
"eternal Universe" is not appropriate in the presence of decreasing modes. If an eternal Universe produces for certain decreasing modes a vanishing value at finite time $\eta_2$, only cosmologies with this property can be extrapolated arbitrarily far to the past. For allowed
generic solutions one only should require that for $\eta_1 \to \infty$ arbitrary initial conditions can be set within the vicinity of a proposed
cosmological solution, and that the evolution for finite $\eta_2$ remains regular for all these initial conditions.

The problem of initial values or the Cauchy problem is time-direction sensitive. Initial conditions have to be set in the past, and evolved
towards the future. Setting "generic initial conditions" at some finite time $\eta_2$, any attempt of arbitrary backwards extrapolation runs
into problems in the presence of strongly decreasing modes. For example, requiring at the end of inflation a ratio of the energy density over the critical density, $\Omega = 1 + \Delta$, with $| \Delta | \sim 10^{-6}$, may at first sight look generic. 
It is well known, however, that for homogeneous cosmological solutions the fraction $\Omega$ moves towards one as time progresses \cite{Starobinsky1980, Guth1981}. If inflation lasts long enough and $\Delta$ is of the order one at the beginning of inflation, the predicted value for $\Delta$ at the end of inflation can be very small. In this case a value $|\Delta|\sim 10^{-6}$ 
contradicts the value predicted for inflationary models with a long duration of inflation. The "initial condition" $|\Delta| \sim 10^{-6}$ at the end of inflation cannot be extrapolated backwards within
the validity of these models. One would not conclude, however, that the models themselves break down. Only $|\Delta| \sim 10^{-6}$ is contradicting 
the prediction of such models.

\section{Quantum scale symmetry}\label{Quantum_scale_symmetry}

The absence of a parameter with dimension of mass or length in the quantum effective action indicates quantum scale symmetry \cite{Wetterich2019}, as characteristic for a fixed point in the renormalization flow of couplings, or more generally, functionals. It is sufficient that there exists one frame where this property is obeyed. Field transformations can introduce a scale, as the Planck mass $M$, in the transformation to the Einstein frame. Such a scale is then a property of the choice of fields, and not of the physical model. The effective action in frames of the type \eqref{S1} is scale invariant if $B$ and $\lambda$ do not involve an intrinsic scale. For the effective action~\eqref{S1}-\eqref{S2A} this is realized in the limit $\chi\to0$ reached at the \qq{beginning} of the universe.

For $\chi>0$ the presence of the scale $\mu$ indicates a breaking of quantum scale symmetry. Such a behavior is characteristic for the vicinity of a fixed point~\cite{Wetterich2019}. The increase of $\chi$ defines a trajectory for the evolution away from a fixed point. Interesting ``crossover cosmologies" arise if an ``UV-fixed point" is realized in the infinite past for $\chi \to 0$, while for increasing $\chi$ the Universe makes a transition to an ``IR-fixed point" in the infinite future for $\chi \to \infty$ \cite{Wetterich2014a, Rubio2017}. This scenario requires that $B(\chi)$ and $\lambda(\chi)$ assume constant values for $\chi \to 0$. 

An UV-fixed point of the renormalization flow corresponds to a scaling solution of the scale dependent effective average action
\begin{equation}\label{87}
\Gamma_k = \int_x \sqrt{g} \left(-\frac{\bar{F}}{2} R + \frac{1}{2} \bar{K} \del^\mu \chi \del_\mu \chi + \bar{U}(\chi)\right),
\end{equation} 
where $k$ is the renormalization scale and the $k$-dependent functions $\bar{F}$, $\bar{K}$ and $\bar{U}$ are given by
\begin{align}\label{88}
\bar{F} = 2\tilde{w}(\tilde{\rho})k^2\ , && \bar{K} = \kappa(\tilde{\rho})\ , && \bar{U} = k^4 u(\tilde{\rho})\ ,\nn\\
\tilde{\rho} = \frac{\rho}{k^2}\ , && \rho=\frac{1}{2}\psi^2\ ,
\end{align}
with $\psi(\chi)$ to be determined below. The scaling form of the effective action for fixed dimensionless fields $\trho$ does not involve any mass scale except the renormalization scale $k$. For $\trho \to 0$ we expand
\begin{align}\label{89}
\tilde{w} = w_0 + \frac{\xi}{2}\trho, && \kappa = \kappa_0 + \kappa_1 \trho, && u = u_0 + \tilde{m}^2 \trho.
\end{align}

We next perform a Weyl scaling to the primordial flat frame, with
\begin{align}\label{91}
w^2 &= \frac{\chi^2}{\bar{F}} = \frac{\chi^2}{k^2(2w_0 + \xi \trho)}.
\end{align}
(The function $w$ in the Weyl scaling should not be confounded with the function $\tilde{w}(\trho)$ in the expansion of $F$. The similarity of the notation is due to historical use.)
In leading order one has
\begin{equation}\label{92}
\chi = \tilde{\mu}(\trho)^\frac{1}{\gamma},
\end{equation}
and finds for the effective action \eqref{S1}
\begin{equation}\label{106}
\lambda = \frac{u_0}{4 w_0^2}\left[1 - \left(\frac{\xi}{2 w_0} - \frac{\tilde{m}^2}{u_0}\right) \left(\frac{\chi}{\tilde{\mu}}\right)^\gamma\right],
\end{equation}
and
\begin{equation}
B = b_1 \left(\frac{\chi}{\tilde{\mu}}\right)^\gamma.
\end{equation}
Here $\gamma$ is given by
\begin{equation}
\gamma = \sqrt{ \frac{4 b_1 w_0}{\kappa_0}},
\end{equation}
with
\begin{align}\label{108}
b_1 = &\frac{6\xi}{w_0}- \frac{6\tilde{m}^2}{u_0} \nonumber \\&+ \frac{9 \kappa_0}{2 w_0}\left(1- \sqrt{1 + \frac{8}{3\kappa_0}\left( \xi - \frac{w_0 \tilde{m}^2}{u_0} \right)}\right).
\end{align}
The derivation of this result can be found in the appendix \ref{Appendix E}.

For the behavior $\chi \to 0$ we observe that for $w_0 > 0$, $\kappa_0 > 0$ the approach of $B$ to zero and of $\lambda$ to $\lambda_0$ now involves a power $(\chi/\tilde{\mu})^\gamma$. This replaces $x(\chi)$ in eq.~\eqref{S2A} by $\trho(\chi)$. We may want to solve the primordial flat frame condition \eqref{eq:FR7} beyond leading order. This amounts to further terms in $B$ and $\lambda$ involving higher order powers of $\trho$. We can again perform a systematic expansion similar to the appendix $B$. Eq.~\eqref{F7} is replaced, however, by
\begin{equation}
\frac{\del \trho}{\del \ln \chi} = \gamma \trho,
\end{equation}
for which the r.h.s vanishes with a different power. We emphasize that the arbitrary mass scale $\tilde{\mu}$ is only introduced by the definition of $\chi$ in eq.~\eqref{92}. It is not a fixed scale of the theory. There is no memory of the renormalization scale $k$ left. Since a scaling solution does not single out a particular scale $k$ this is a natural outcome. The same feature is observed if one transforms to the Einstein frame. In this case $M$ is introduced by the Weyl transformation, and no memory of $k$ is left either.

Scaling solutions for quantum gravity coupled to a scalar field have been computed by functional renormalization. For computations of the overall form of the potential $u(\trho)$ and the curvature coefficient $\tilde{w}(\trho)$ see refs.~\cite{HPRW,HPW2,PRW,CWMY,CWESNP}. Once remaining open points in these computations are clarified, the function $\kappa(\trho)$ is computed, and the behavior of higher derivative terms is established, such computations could lead to quantum gravity predictions for properties of inflation.

\section{Lightlike vacuum in the Einstein frame}\label{Lightlike_vacuum_in_the_Einstein_frame}

The physical properties of the lightlike vacuum can be seen in a rather direct way in the primordial flat frame. All particle masses vanish in the infinite past since they are proportional to $\chi$ and $\chi$ goes to zero. The homogeneous Universe can be extended to the infinite past.
This extends to neighboring inhomogeneous cosmologies provided that the propagators remain finite for $\chi \to 0$. Cosmic time $t$ and conformal time $\eta$ are proportional to each other for Minkowski space, and for both times the regular cosmological solution can be extended to infinite negative values. Physical properties should be independent of the choice of fields used to parameterize them. The physical properties should therefore be the same in the Einstein frame. With a discussion of observable particle masses and physical time in the Einstein frame we will show that this is indeed the case.

\subsection*{Lightlike vacuum}

While the property of emptiness of the universe in the inflationary stage is well known, the lightlike behavior of excitations needs a more detailed discussion \cite{Wetterich2014}. We aim here for physical properties that are at least in principle observable by a gedankenexperiment. In a quantum field theory observable quantities should not depend on the choice of fields used to describe them. In more technical terms they should not depend on the ``frame" used for the metric field. Observable quantities have to be dimensionless. In our case the relevant dimensionless quantity is the ratio mass over momentum $m/p$. 
It is this ratio that matters for the distinction between relativistic and non-relativistic particles and for the issue of the use of proper
time for a physical time definition. The ratio between particle mass and Planck mass, which determines the strength of gravity, is not
relevant for this purpose.
For $m/p \to 0$ the particle becomes ultrarelativistic and propagates like light -- the difference to the propagation of a photon disappears in this limit. 

In the familiar Einstein frame with fixed particle mass $m$ the universe expands roughly exponentially during the inflationary epoch,
\begin{equation}\label{exp inflation}
a(t) \approx \exp\{ H_E (t-t_i) \}a(t_i),
\end{equation}
with $a(t)$ the scale factor in the Robertson-Walker metric and $t$ cosmic time. As a consequence the physical momentum of a particle, $p = k/a$, with $k$ the comoving momentum, decreases exponentially,
\begin{equation}
p(t) \approx \exp\{-H_E (t-t_i)\} p(t_i).
\end{equation}
A slow time dependence of $H_E$ does not change the situation.

Consider now at the time $t_0$ at the end of inflation a superheavy particle with mass $m$ of the order of the Planck mass and a very small momentum, say $p(t_0) = 10^{-10}m$. At this time the particle is non-relativistic, $m/p = 10^{10}$, and has a momentum much smaller than the expansion rate of the Universe $H_E$. Looking at a time $t$ sixty $e$-folds before the end of inflation, one finds already a rather small ratio $(m/p) \approx e^{-60} 10^{10} \approx 10^{-16}$, and the particle is ultrarelativistic. Going back further the ratio further decreases rapidly. For nucleons with the same momentum the ratio is a factor $10^{-18}$ smaller at any time. We can repeat the argument by placing $t_i$ at some arbitrary moment during inflation. If inflation lasts long enough before $t_i$ the particle will again be ultrarelativistic at sufficiently early $t$. 

In particular, if the inflationary epoch has no ``beginning event" at some $\bar{t}$, any nonzero momentum $p(t_i)$ will diverge as $t$ goes to minus infinity. A more detailed discussion would consider the evolution of momentum distributions, but the sense in which we speak about a ``lightlike" vacuum should already be clear: towards the beginning particles propagate similar to photons. (There is always a tail of extremely small momenta $p(t_i)$ for which $(m/p)(t)$ remains larger than one at any given finite $t$. See ref. \cite{Wetterich2014} for a discussion of different limits.) We will focus here on inflationary scenarios without a ``beginning event" and discuss alternatives at the end of this note. Towards the ``beginning" $t \to -\infty$ all particles then become massless in physical terms, justifying the notion of a lightlike vacuum.
Indeed, in terms of the relevant ratio $m/p$ all particles become massless towards the beginning, irrespective of the choice of frame. Only the particular picture or "mechanism" for the realisation of massless particles differs
between the frames. In the primordial flat frame it is due to a field dependence of the masses and $\chi \to 0$, while in the Einstein
frame it arises from a vanishing scale factor $a_E \to 0$.

Massless particles are an indication of the possibility of unbroken scale symmetry.
In the scaling frame or primordial flat frame scale symmetry is directly visible. In the Einstein frame it is obscured by the introduction of a
fixed mass scale $M$ in the Weyl transformation.

\subsection*{Physical time}

The statement that the beginning epoch of the Universe can last since ever may encounter more doubts. In the Einstein frame one observes that a vanishing scale factor is reached for infinite negative cosmic time in one class of inflationary models comprising, for example, Starobinsky inflation. In another class a vanishing scale factor may be reached at finite $t$. It has been argued that the limit $a \to 0$ corresponds to a singularity \cite{Penrose1965, Hawking1966}, since geodesics are not complete. This geodesic incompleteness has been established under rather general conditions of a positive average expansion rate~\cite{ Borde2001, Mithani2012}. This ``geodesic incompleteness" has given rise to the opinion that in standard inflationary cosmology the Universe starts with a singularity and that the inflationary epoch only lasts for an extremely short physical time, say $10^{-40}$ seconds. Many alternative beginnings have been proposed that aim to avoid this ``initial singularity".

In the primordial flat frame the geometry is geodesically complete, being simply Minkowski space. Since field transformations of the metric change the geometry, it is obvious that purely geometric properties do not correspond to observable quantities. One has to concentrate on observable quantities that are necessarily independent of the chosen metric frame and to understand how they behave in the Einstein frame.

We argue here that in standard models of inflation there may be no physical initial singularity. For a suitable definition physical time $t_{ph}$ extends to the infinite past, $t_{ph} \to -\infty$. We want to show here how infinite physical time is found in the familiar Einstein frame with fixed particle masses. We note that proper time is useful for many purposes, but it is not a reasonable physical time when we consider the Universe towards its ``beginning". As is well known, proper time cannot be used for massless particles, and we have just seen that all particles become effectively massless for $\eta \to -\infty$. Furthermore, proper time is not a frame invariant quantity but rather depends on the specific choice of a metric field. A detailed discussion \cite{Wetterich2014} reveals that proper time is indeed inappropriate for the limit $a \to 0$.

A frame invariant quantity multiplies the infinitesimal proper time element by the particle mass \cite{BAST, Wetterich2014},\bel{PTA}
\text{d}\tilde\tau=m\text{d}\tau\ .
\ee
If the integral $\tilde\tau$ of this quantitiy remains finite, this corresponds to geodesic incompleteness in the Einstein frame \cite{Wetterich2014}. The multiplicative factor of the particle mass makes the limitations of this quantity apparent. Even for a Minkowski geometry the integral can be finite if the particle mass goes to zero sufficiently fast. This is another facet of the statement that proper time becomes problematic if effective physical particle masses go to zero. This simple observation also reveals that the geometric notion of geodesic completeness is not appropriate for a discussion of physical singularities. One would not call a Minkowski geometry incomplete because $\tilde\tau$ remains finite for $t\to-\infty$ for a particle whose mass vanishes sufficiently fast for $t\to-\infty$. On the other hand, finite $\tilde\tau\gl t\to-\infty\gr$ evaluated on a physical particle trajectory may be considered as problematic.

We propose to define physical time by some type of counting the number of oscillations. This is close to what is done for the present practical time definition by counting the number of oscillations for a given transition in an atomic clock. There will be no atomic clocks in the inflationary epoch, but wave functions of particles, mode functions or propagators still show oscillatory behavior. We may use the number of oscillations of the wave function for photons for a given comoving momentum $k$. Equivalently, one could use gravitational waves or the wave functions for gravitons. Physical time is then proportional to the number of oscillations of this \qq{photon clock}.

Expressed in conformal time $\eta$ the wave functions of massless particles in an isotropic homogeneous Universe obeys the evolution equation
\begin{equation}
(\partial_\eta^2 + 2 H a \partial_\eta + k^2) \varphi_k = 0.
\end{equation}
Here the complex functions $\varphi_k(\eta)$ are Fourier modes, corresponding to eigenstates of comoving momentum $k$, $H = \partial_t a/a,$
$\mathscr{H} = Ha = \partial_\eta \ln(a),$ $a d\eta = dt$. We factor out the Hubble damping,
\begin{align}
\tilde{\varphi_k} = a \varphi_k, && \left( \partial_\eta^2 + k^2 - \frac{a^2R}{6}\right) \tilde{\varphi_k} = 0,
\end{align}
with $R$ the curvature scalar. In the beginning of inflation one has $|a^2 R| \ll k^2$ for any $k$. Then the number of oscillations $n_k$ is proportional to conformal time
\begin{equation}\label{PT3}
n_k = \frac{k \eta}{2\pi}.
\end{equation}

We define a dimensionless physical time by the number of oscillations $n_k$ that a photon wave function undergoes starting from some reference point. If we choose the end of inflation as a reference point, physical time is negative during inflation. Different modes $k$ define different clocks. The corresponding physical time defined by the counting of different clocks is directly related by its proportionality to $k$. As done for atomic clocks, one selects a ``reference clock" by some reference comoving wavelength $\sim k^{-1}$ and gauges the other clocks correspondingly. Also the oscillations of the wave functions of massive particles at rest in the cosmic reference frame can be gauged to the "reference photon clock". For $k \neq 0$ one has a finite number of oscillations of the particle wave function for one oscillation of the reference photon clock.

Due to the simple relation \eqref{PT3} we propose to use conformal time $\eta$ as a good proxy for physical time. It is directly proportional to physical time for all geometries for which $R$ can be neglected as compared to the squared inverse wavelength or squared physical momentum $q^2 = k^2/a^2$. Conformal time introduces units of time $\sim k^{-1}$. It guarantees the ``gauging of clocks" with different wavelength. As an important aspect for our discussion, conformal time is invariant under Weyl scalings of the metric. It is the same in all frames related by conformal transformations.

Physical time given by the number of photon oscillations remains well defined for inhomogeneous cosmologies. For any given length scale of the inhomogeneities one can consider photons with wavelength much shorter than this scale. The number of oscillations of the photon wave function will only be affected very mildly by a weak inhomogeneity. As long as we consider only weak inhomogeneities we may take the conformal time of the neighboring homogeneous cosmology as a good proxy for physical time even in the presence of inhomogeneities.

Physical ``oscillation time" is precisely the same in all frames. The counting of oscillations is discrete and therefore independent of the choice of fields. It is not affected by coordinate transformations. The universality of conformal time is more restricted. Its equivalence with oscillation time holds for homogeneous isotropic cosmologies if the curvature scalar is negligible and if the particles are massless. For our purpose this is sufficient and we will use conformal time as a definition for physical time here.

Measured in oscillation time the time distance to the ``big bang singularity" at ``$a = 0$" is infinite. The clocks tick an infinite number of times. For inflationary models without a beginning event the Universe is eternal, it has existed forever in physical time. In the Einstein frame the cosmic time interval $\Delta t$ between two ticks gets shorter and shorter as one approaches $a \to 0$, whereas in the primordial flat frame $\Delta t$ goes to a constant. The number of ticks is the same. Both conformal time and physical oscillation time can be extrapolated backwards to the infinite past $\eta \to - \infty$. With conformal time being the same in both frames, it is the mapping to cosmic time $t(\eta)$ that depends on the frame. While in the primordial flat frame the ratio of intervals between two ticks,
\begin{equation}
\frac{\Delta t}{\Delta \eta} \approx a(\eta)
\end{equation}  
is given by the constant $\bar{a}$, it goes to zero in the Einstein frame. This is the reason why in the Einstein frame the cosmic time elapsed since the ``big bang singularity" can be finite, $t_{BB}\approx 13.7$ billion years, despite oscillation time and conformal time being infinite.

Expressed in conformal time the history of the hot big bang Universe and inflation looks less dramatic. Measured in physical (conformal) time the ``conformal age" of the Universe since the end of inflation amounts to around 46 billion years. For a cosmological epoch where
\begin{equation}
\frac{a(t)}{a(t_{in})} = \left( \frac{t}{t_{in}} \right)^{\frac{2}{n}},
\end{equation}
with $n = 3(4)$ for matter (radiation) domination, one finds
\begin{equation}
\eta(t_1) -\eta (t_2) = \left( 1- \frac{2}{n} \right)^{-1} t_{in}^{\frac{2}{n}} a_{in}^{-1} \left( t_1^{1-\frac{2}{n}} - t_2^{1-\frac{2}{n}} \right).
\end{equation}
For some time $t$ in the radiation dominated epoch one has
\begin{equation}
\eta_{eq} - \eta(t) = 2 z_{eq} \lb(t_{eq} - t_{eq}^{\frac{1}{2}} t^{\frac{1}{2}} \rb).
\end{equation}
The difference in conformal time is much larger than the difference $t_{eq} -t$ in cosmic time, being enhanced by the redshift $2 z_{eq} \approx 7000$ for matter radiation equality. In physical time the radiation dominated epoch between the end of inflation and matter-radiation equality lasts for $3.3\cdot 10^8$yr or around one percent of the conformal age of the Universe since the end of inflation.

The most important qualitative difference between physical time and cosmic time $t$ occurs for the (almost) exponential expansion (\ref{exp inflation}) during inflation. For $t_i = t_0$ the end of inflation and $t$ some time during inflation one has
\begin{align}\label{eq:PT7}
\eta(t_0) - \eta(t) &= \frac{1}{H_E} \lb( \frac{1}{a(t)} - \frac{1}{a(t_0)}\rb) \nonumber \\
&= \frac{1}{H_E a(t_0)} \lb(\exp\{H_E(t_0-t)\}-1\rb).
\end{align}
Physical time diverges for  $a(t) \to 0$, and the Universe lasts since ever when time is measured in physical units.

One may also map oscillation time to proper time. Proper time is not useful for the limit $a \to 0$ since all particles become massless. Nevertheless, it can become a useful concept for later stages in the evolution of the Universe. Proper time is not independent of the choice of frames. For its relation to oscillating wave functions for massive particles, see ref.~\cite{Wetterich2014}.
Furthermore, we can compute $\tilde\tau(\eta)$ for the frame invariant integrated products of proper time intervals and particle mass. In the Einstein frame $\tilde\tau$ is proportional to proper time $\tau$ for a massive particle, while in the primordial flat frame one has
\bel{PTB}
\text{d}\tilde\tau\sim\chi(\eta)\text{d}\tau(\eta)=\chi(\eta)\frac{\text{d}\tau}{\text{d}\eta}\text{d}\eta\
\ee
with $\text{d}\tau/\text{d}\eta$ evaluated on the trajectory of the massive particle which differs from the timelike geodesics~\cite{Wetterich2014}.

For a particle at rest $\tilde\tau(\eta\to-\infty)$ diverges for many models. With asymptotically constant $\text{d}\tau/\text{d}\eta$ the scalar field vanishes slowly enough, cf. eq.~\eqref{AF1}. On the other hand, for a free particle with non-zero comoving momentum $\tilde\tau$ remains finite, due to finite proper time in the Einstein frame or due to the combination with $\chi(\eta\to-\infty)\to0$ in the primordial flat frame. This issue is the frame-invariant formulation of the incomplete timelike geodesics in the Einstein frame. Free particles with non-zero momentum have to start with an infinite ratio of momentum over mass in the beginning for $\eta\to-\infty$. It is not obvious if one can conceive physical processes for which the number of oscillations is proportional to $\tilde\tau$. Such a clock would start with a non-oscillating behavior for $\eta\to-\infty$. It seems much easier to base the concept of physical time on the photon clock, and to map this clock only in later stages of the evolution to proper time of massive particles or to $\tilde\tau$. Finite $\tilde\tau(\eta\to-\infty)$ does not mean that time stops, but rather that a possible associated clock becomes inappropriate since the underlying physical process is not oscillating in this limit.

\section{Discussion}\label{Discussion}

In this paper we discuss the inflationary epoch in the early stages of the evolution of the Universe. We describe standard inflationary models in the primordial flat frame for which geometry approaches Minkowski space as physical time goes to the infinite past. For several  models, as Starobinsky inflation or chaotic inflation, we construct the primordial flat frame explicitly. In this frame the effective action for the metric and a scalar field features a dynamical Planck mass given by the scalar field and a negative coefficient of the scalar kinetic term. Nevertheless, these models of variable gravity are stable. We solve the field equations explicitly and show that geometry approaches flat space for cosmic time $t \to -\infty$.

As a consequence, the Universe can exist since an infinite time in these standard models of inflation. The Universe can be eternal in the past and the future. This does not necessarily imply that our Universe has been forever in this ``beginning stage" of inflation. This is one possibility, but other possibilities as a bounce crossing the big bang singularity \cite{STTU,Bars2013,Kamenshchik2016,BRAP, CCRK, CWCBB}, or creation by a fluctuation in a finite region of a multiverse \cite{Linde1983,Shafi1983}, remain possible as well. In this case the solutions discussed in this paper may have been approached at a finite time $t$, while the limit $t \to - \infty$ is different. Many aspects of our discussion apply to such scenarios as well.

The question arises if our Universe emerges in the infinite past from a singularity or from a regular solution. In this context one has to
differentiate between field singularities and "physical singularities". Field singularities are an artifact of a particular choice of 
"field coordinates" in field space. For cosmology, this concerns the choice of the metric frame. For the absence of a physical singularity
it is sufficient that one frame exists for which one can establish that all "observable quantities" remain finite. A field transformation 
to another frame may be singular at certain points in field space. In this case it introduces a field singularity. This is what happens 
for the homogeneous cosmologies discussed in this paper. In the primordial flat frame everything remains regular. Singularities in the Einstein frame arise from the Weyl transformation becoming singular for a vanishing scalar field. The situation is  analogous to regular and
singular coordinate patches in geometry.

For the issue of a possible physical singularity one has to differentiate between the homogeneous isotropic solution and neighboring inhomogeneous 
cosmologies. The homogeneous solution shows no singularity. With the present approximation to the effective action, however, the standard
inflationary models lead to physical singularities if arbitrary neighboring inhomogeneous solutions are extrapolated backwards to the infinite past. Even though this
singularity appears only in the infinite past in physical time, we should understand its precise status and implications.

One of the aims of the present paper is to shed new light on the status of these singularities. A first important observation states that
not all inhomogeneous solutions need a regular continuation backwards to the infinite past. Part of the inhomogeneous cosmological solutions
can, and often will, diverge if extrapolated backwards without signaling a physical problem. These singularities may even occur at 
finite physical time. The reason is spontaneous breaking of the time reversal symmetry by the homogeneous cosmological solution. As a result
the evolution of small fluctuations around this solution involves damping. If for certain "decreasing fluctuation modes" the damping is strong
enough, their amplitude is predicted to be bounded within a certain interval at some given time in cosmology, say the end of inflation. 
Only values (or "initial conditions") within this allowed interval can be extrapolated consistently to the past. Initial conditions outside
this interval contradict the prediction of the model and are not allowed. 

In view of this arrow of time arbitrary initial conditions should only be set in the infinite past if one wants to study the issue of physical
singularities. In many circumstances it is predicted that for generic initial conditions in a certain range the amplitude of certain
decreasing modes is zero at the end of inflation. Non-zero values for inhomogeneities in these modes are then expected to lead to a singular
behavior if one extrapolates backwards. If the prediction of a zero value concerns whole modes, a generic inhomogeneous cosmology at the end
of inflation cannot be extrapolated backwards to the infinite past. These types of \qq{backward singularities} are a necessary consequence of modes damped to zero and should not be regarded as a physical problem.

The absence of inhomogeneities in the decreasing modes is not the central problem for the issue of possible physical singularities. The problem
concerns the observable fluctuations. Their mean amplitude can be computed from the second functional derivative of the effective action and
therefore involves the propagators for the physical fluctuations. These fluctuations are observed to be different from zero. As a consequence, 
one should be able to extrapolate the inhomogeneous cosmological solutions corresponding to these modes backwards in time and use them for the investigation
of physical singularities for a given cosmological model. In the present approximations or models the relative amplitude of these observable modes 
diverges as one approaches the infinite past, This indicates the breakdown of the linear approximation for small relative inhomogeneities.

The divergence concerns only the relative inhomogeneous fluctuations. We find that the associated frame-invariant metric for these fluctuations remains finite in the infinite past. The divergence arises since the frame-invariant homogeneous average metric vanishes faster for $\eta\to-\infty$ than the inhomogeneous counterpart. The universe begins inhomogeneous, but not singular. As time increases, the homogeneous part grows faster than the inhomogeneous part, and relative inhomogeneities become small. This describes well the almost homogeneous universe at the end of the inflationary epoch. Even though the linear approximation for inhomogeneities is no longer valid at the \qq{beginning} of the universe we believe that the overall picture remains valid beyond the linear approximation. Initial conditions should be set far in the past. One may then verify, for example by numerical simulations, that our picture holds as time increases.

The amplitude of the observable fluctuations 
is directly linked to the propagator for these modes. In the approximation \eqref{S1} the propagator for the relative graviton fluctuations diverges for $\chi \to 0$.
The inverse propagator is given by the second functional derivative of the effective action with respect to the graviton fluctuations. It vanishes for
a vanishing scalar field $\chi$. Since $\chi$ vanishes in the infinite past, the graviton propagator has to diverge in the infinite past. In the primordial flat frame this produces a singularity for the corresponding inhomogeneous metric, but not for the frame-invariant metric.

In the primordial flat frame it becomes apparent how divergent relative fluctuations could possibly be avoided. One needs a modification of the graviton
propagator (and possibly also other propagators) for momenta that are large as compared to $\chi$. (In the Einstein frame this concerns momenta
larger than the Planck mass.) In any quantum field theory for the metric one indeeds expects that the quantum effective action contains terms 
involving higher derivatives of the metric. The one contributing to the graviton propagator in flat space is given by eq.~\eqref{IH23}. If a 
consistent function $D(q^2)$ can be found, the divergence of the relative inhomogeneities will disappear whenever $D$ differs from zero for non-zero momenta. In this case
the graviton propagator for $k \neq 0$ remains finite in momentum space, and finiteness for the Fourier transform from frequency to $\eta$ can be expected.
If the graviton propagator remains finite for $\eta \to -\infty$, no divergence will occur in this sector of inhomogeneous relative fluctuations.

Finding an acceptable form of $D(q^2)$ is a central issue in all approaches to quantum gravity. This term governs the behavior for large
momenta and therefore the short distance behavior of gravity in the traceless transversal tensor (graviton) sector. A simple 
polynomial form produces ghosts and is most likely unacceptable. An acceptable form is not yet known. For any consistent quantum field 
theory for the metric it has to exist. If not, this indicates that the short distance behavior of gravity has to be formulated in terms
of degrees of freedom different from the metric.

Fortunately, these issues concern only the very early stages of inflationary cosmology. We have determined the range of $k$ and $\eta$ for which
the higher derivative terms of the type \eqref{IH23} become relevant. This is typically outside the range relevant for the observed fluctuation 
spectrum.

The main theme of the present work concerns the formulation of standard inflationary models in the primordial flat frame.
While a discussion in a different metric frame can always be done, one may ask if such a reformulation is useful. We may list several points for which the formulation in the primordial flat frame leads to new insights.

(1) The homogeneous isotropic solution in the primordial flat frame shows no singularity. It is regular for all $t$, including the limit $t \to -\infty$. The metric, the inverse metric and the scalar field remain finite.

(2) The solutions approach Minkowski space for $t \to - \infty$. This makes the interpretation of time rather simple. Conformal time is proportional to cosmic time and both can be continued to infinite values in the past.

(3) The fluctuation problem, based on the propagator and mode functions, becomes rather simple in flat space. It is sufficient to establish the propagator in Minkowski space, taking into account a time varying scalar field.

(4) The role of higher derivative terms in the effective action becomes comparatively simple.
Unless their coefficients diverge they play no role for the homogeneous isotropic solution in the limit $t \to - \infty$. The squared Weyl tensor does not modify the homogeneous isotropic field equations, and the squared curvature tensor $R^2$ becomes subleading for the solutions found, with $R^2 / (\chi^2 R) \to 0$ for $t \to - \infty$. On the other hand, the higher derivative terms will dominate the graviton propagator at non-zero momentum in the limit $\chi \to 0$, unless the coefficient function $D$ for the squared Weyl tensor vanishes.

(5) It is directly visible that all particles are massless for $t\to - \infty$. This follows from the solution $\chi(t\to -\infty) \to 0$, given that for variable gravity in the primordial flat frame all particle masses are proportional to $\chi$.

(6) Quantum scale symmetry is directly visible for $t \to - \infty$. Indeed, for $\chi \to 0$ the effective action in the primordial flat frame does not involve any parameter with dimension of mass or length. The choice of fields for the primordial flat frame is one for which the scale transformations are implemented in a simple manner.

(7) The quantum scale symmetry of the effective action allows for a direct connection to an ultraviolet fixed point for quantum gravity, as postulated for asymptotic safety \cite{ASW,ASR}. Fixpoints induce exact quantum scale symmetry \cite{Wetterich2019}.

(8) From the point of view of the functional renormalization flow the choice of fields used for the primordial flat frame minimizes the dependence of the scaling solution on the renormalization scale $k$. For the scaling solution the dimensionless functions in the (average) effective action only depend on $\chi^2/k^2$. If the limits $\chi \to 0$ and $\chi \to \infty$ do not diverge, the effective action becomes independent of $k$ in these limits.

One may also ask what one can learn from the comparison of different frames. The most important point in this respect is the focus on physical properties. They are formulated in the form of quantities that are, in principle, observable, at least by a ``Gedankenexperiment". For the beginning epoch or inflationary epoch key properties are:

(i) The state of the Universe is a vacuum, characterized by expectation values of fields and fluctuations, as encoded in propagators. In the early stages propagating particles are extremely rare, and there is almost no entropy. Particles and entropy are created during the heating period after the end of inflation.

(ii) All particles are almost massless, and precisely massless in the infinite past. This property does not only concern the relativistic propagation of the rare particles, but also the relativistic form of the propagator and the associated primordial fluctuation spectrum.

(iii) The evolution in the early stages of the Universe is very slow in physical time. For the primordial flat frame both the relative change of the scalar field $\dot \chi/\chi$, and the relative change of the scale factor, $H = \dot a/a$, go to zero for $t \to -\infty$. They define the inverse characteristic time scales. The change in geometry is even much slower than the change in the scalar field, $H \chi/\dot \chi \to 0$ for $t \to - \infty$.

The physical properties of the inflationary epoch are the lightlike vacuum. In contrast to the view of a very tumultuous epoch we may call this stage the ``great emptiness".

\bigskip

Acknowledgment: The author would like to thank V. Rubakov for many profound discussions about physical time.

\appendix
\titleformat{\section}[display]{\normalfont\bfseries\large}{\appendixname{} \thesection.}{0pt}{\bfseries\large}

\section{Cosmological solutions for the  primordial flat frame for Starobinsky inflation}\label{Appendix A}

In this appendix we discuss the field equations and their solution for the model \eqref{S1},\eqref{S2}. 
The metric field equations derived from the action \eqref{S1} read for a Robertson-Walker metric $(R = 12 H^2 + 6\dot H)$
\begin{equation}\label{eq:FR9}
3 \chi^2 H^2 = \lambda \chi^4 + \frac{B-6}{2} \dot \chi^2 - 6H \chi \dot \chi,
\end{equation}
\begin{equation}\label{eq:FR10}
\chi^2 R = 4 \lambda \chi^4 - B\dot \chi^2 - 6 \chi ( \ddot \chi  + 3 H \dot \chi),
\end{equation}
and the scalar field equation is given by
\begin{equation}\label{eq:FR8}
(B-6)(\ddot \chi + 3H \dot \chi) = - 4 \lambda \chi^3 - \chi^4 \frac{\del \lambda}{\del \chi} + \chi R - \frac{1}{2} \frac{\del B }{\del \chi} \dot \chi^2.
\end{equation}

Our aim is the investigation of possible solutions of the field equations \eqref{eq:FR9}-\eqref{eq:FR8} with a geometry close to flat space. For this purpose we first bring them into a suitable form.
For $\chi \neq 0$ we can insert eq.~\eqref{eq:FR10} into eq.~\eqref{eq:FR8},
\begin{equation}\label{F1}
\ddot{\chi} + 3 H \dot{\chi} = - \frac{\del \lambda}{\del \ln \chi} \frac{\chi^3}{B} - \left(1 + \frac{1}{2} \frac{\del \ln B}{\del \ln \chi}\right) \frac{\dot{\chi}^2}{\chi}.
\end{equation}
Taking linear combinations, the two other field equations become
\begin{align}\label{F2}
H^2 + 2 H \frac{\dot{\chi}}{\chi} = f_1, && f_1 = \frac{\lambda}{3} \chi^2 + \frac{B-6}{6} \frac{\dot{\chi}^2}{\chi^2}.
\end{align}
and
\begin{align}\label{F3}
&\dot{H} - 4H \frac{\dot{\chi}}{\chi} = f_2, \nonumber\\
&f_2 = \frac{\del \lambda}{\del \ln \chi} \frac{\chi^2}{B} + \frac{1}{2} \left( 6 - B + \frac{\del \ln B}{\del \ln \chi}\right) \frac{\dot{\chi}^2}{\chi^2}.
\end{align}
A similar form is found for eq.~\eqref{F1},
\begin{align}\label{F4}
&3 H \frac{\dot{\chi}}{\chi} = f_3, \nonumber\\
& f_3 = - \frac{\del \lambda}{\del \ln \chi} \frac{\chi^2}{B} - \left(1 + \frac{1}{2}  \frac{\del \ln B}{\del \ln \chi}\right) \frac{\dot{\chi}^2}{\chi^2} - \frac{\ddot{\chi}}{\chi}.
\end{align}
Exact flat space solutions require $f_1 = f_2 = f_3 = 0$.

For the solution of the field equations \eqref{F2}-\eqref{F4} for the type of models considered here we make the ansatz $(\chi \neq 0)$
\begin{align}\label{S3}
\frac{\dot{\chi}^2}{\chi^4} = \tilde{c}^2(\chi), \quad \dot{\chi} = \tilde{c} \chi^2,
\end{align}
which implies
\begin{equation}
\frac{\ddot{\chi}}{\chi} = \tilde{c}^2\left(2 + \frac{\del \ln \tilde{c}}{\del \ln \chi}\right) \chi^2.
\end{equation}
The r.h.s. of eqs.~\eqref{F2}-\eqref{F4} become
\begin{equation}\label{S5}
f_k = h_k \chi^2,
\end{equation}
with
\begin{align}
h_1 &= \frac{\lambda}{3} - \lb(1- \frac{B}{6}\rb) \tilde{c}^2, \\
h_2 &= \frac{1}{B} \frac{\del\lambda}{\del \ln \chi} + \frac{1}{2}\lb(6 - B + \frac{\del \ln B}{\del \ln \chi}\rb) \tilde{c}^2, \\
h_3 &= - \frac{1}{B} \frac{\del \lambda}{\del \ln \chi} - \frac{1}{2}\lb(6+ \frac{\del \ln B}{\del \ln \chi} + \frac{\del \ln (\tilde{c}^2)}{\del \ln \chi}\rb)\tilde{c}^2.\label{S6C}
\end{align}

The functions $h_k$ and the function $\tilde{c}$ in eq.~\eqref{S3} have to obey certain relations which are required by the self-consistency of the field equations. We observe the relations
\begin{align}\label{S7A}
\frac{H}{\chi} = \frac{h_3}{3 \tilde{c}}, \quad \frac{H^2}{\chi^2} = h_1 - \frac{2}{3}h_3.
\end{align}
The consistency of the system of field equations therefore requires
\begin{align}\label{S7B}
\frac{h_3^2}{9 \tilde{c}^2} = h_1 - \frac{2}{3} h_3, \quad \tilde{c}^2 = \frac{h_3^2}{9 h_1 - 6 h_3}.
\end{align}
Eq.~\eqref{S7B} is a first consistency relation.

From the first equation \eqref{S7A} we infer
\begin{align}
\dot{H} &= \left[ \frac{h_3}{3\tilde{c}} + \frac{\del}{\del \ln \chi} \lb( \frac{h_3}{3\tilde{c}} \rb) \right] \dot{\chi} \nonumber\\
&= \frac{h_3}{3} \lb(1 + \frac{\del \ln h_3}{\del \ln \chi} - \frac{\del \ln \tilde{c}}{\del \ln 
\chi} \rb) \chi^2.
\end{align} 
Eqs.~\eqref{F3}, \eqref{F4} therefore impose the relation
\begin{equation}\label{S7D}
h_2 + \frac{4}{3}h_3 = \frac{h_3}{3} \lb( 1 + \frac{\del\ln h_3}{\del \ln \chi} - \frac{\del \ln \tilde{c}}{\del \ln \chi} \rb),
\end{equation}
or
\begin{equation}\label{S7E}
h_2 + h_3 = \frac{1}{3} \frac{\del h_3}{\del \ln \chi} - \frac{h_3}{6} \frac{\del \ln(\tilde{c}^2)}{\del \ln \chi}.
\end{equation}
Eq.~\eqref{S7E} is the second consistency relation. 

Solutions with $H=0$, $\dot{H} = 0$ are approached for limits for which all $h_k \chi^2$ vanish. We are interested in solutions for which $\chi$ vanishes. The approach to a constant scale factor will require that the functions $h_k$ vanish fast enough for $\chi \to 0$. We consider models for which $B$ vanishes in this limit, as for the model \eqref{S2}. Then the coefficient $h_1$ vanishes if for $\chi \to 0$ or $x \to 0$ one has
\begin{equation}\label{S8}
\tilde{c}^2 = \frac{\lambda}{3}.
\end{equation}
The coefficients $h_2$ and $h_3$ also vanish in leading order if
\begin{equation}
\frac{1}{B} \frac{\del \lambda}{\del\ln\chi} = - 3 \tilde{c}^2.
\end{equation}
Thus all $h_k$ go to zero if besides eq.~\eqref{S8} one has
\begin{equation}\label{S10}
\frac{\del\ln\lambda}{\del\ln\chi} = -B.
\end{equation}

For the model \eqref{S2} we are interested in $\chi^2 \ll \mu^2/c_t$ such that the term $c_t$ in the logarithm \eqref{S2} can be neglected, and
\begin{equation}
\frac{\del x}{\del \ln \chi} = 2 x^2.
\end{equation}
For the model \eqref{S2} one obtains 
\begin{equation}
\frac{\del \ln \lambda}{\del \ln \chi} = -6x^2 \frac{1-\frac{5x}{3}\ln \lb( \frac{2}{3x}\rb) + \frac{5x}{6}}{1- \frac{3x}{2} + \frac{5x^2}{4} \ln \lb(  \frac{2}{3x}\rb)},
\end{equation}
and
\begin{equation}
\frac{\del B}{\del \ln \chi} = 24x^3 \lb( 1-\frac{x}{2}\ln \lb( \frac{2}{3x}\rb) + \frac{10x}{3}\rb).
\end{equation}
For $x\to 0$ eq.~\eqref{S10} is indeed obeyed. Since $\lambda$ approaches a constant $\lambda_0$, $\tilde{c}$ also goes to a constant. The solution of eq.~\eqref{S3} for the scalar field reads
\begin{equation}\label{S14}
\frac{1}{\chi} = \tilde{c} (t_0 - t) + \delta,
\end{equation}
where $\delta$ remains bounded for $t\to - \infty$. In the infinite past for $t\to -\infty$ the scalar field vanishes $\sim [\tilde{c}(t_0 - t)]^{-1}$.

For a more detailed investigation of the solution for $t\to -\infty$ we parameterize
\begin{equation}\label{S15}
\tilde{c}^2 = \frac{\lambda}{3} (1+\Delta).
\end{equation}
This results in
\begin{equation}
\frac{\ddot{\chi}}{\chi^3} = \lb(2+ \frac{1}{2} \frac{\del \ln \lambda}{\del \ln \chi}\rb) \tilde{c}^2 + \frac{\lambda \dot{\Delta}}{6 \tilde{c} \chi},
\end{equation}
replacing in the equation \eqref{S6C} for $h_3$
\begin{equation}\label{S17}
\frac{\del\ln\tilde{c}}{\del \ln \chi} \to \frac{1}{2} \frac{\del\ln\lambda}{\del \ln\chi} + \frac{\lambda \dot{\Delta}}{6 \tilde{c}^3 \chi}.
\end{equation}
If $\Delta$ is a function of $\chi$ one has
\begin{equation}\label{S17A}
\frac{\del_t\Delta}{\chi} = \frac{1}{\chi^2} \frac{\del \Delta}{\del \ln\chi} \dot{\chi} = \tilde{c} \frac{\del \Delta}{\del \ln \chi},
\end{equation}
and therefore
\begin{equation}
\frac{\lambda \del_t \Delta}{6 \tilde{c}^3 \chi} = \frac{1}{2} \frac{\del \ln (1+\Delta)}{\del \ln \chi}.
\end{equation}
We will find $\Delta \ll 1$, such that the second term $\sim \del_t \Delta$ in eq.~\eqref{S17} can be neglected in leading order.

Let us first look at the coefficients $h_k$ for $\Delta = 0$. They are all proportional to $\lambda$
\begin{equation}
h_1 = \frac{\lambda B}{18},
\end{equation}
\begin{align}
h_2 &= \frac{\lambda}{B} \lb( \frac{\del \ln \lambda}{\del \ln \chi} + B - \frac{B^2}{6} + \frac{1}{6} \frac{\del B}{\del \ln \chi} \rb) \nonumber \\
&= \frac{41 x^4 \lambda}{6 B}\lb( 3\ln \lb( \frac{2}{3x}\rb) -2 \rb), 
\end{align}
\begin{align}
h_2 + h_3 &= - \frac{\tilde{c}^2}{2}\lb(B + \frac{\del \ln \lambda}{\del \ln \chi}\rb) \nonumber\\
&= \lambda x^3 \lb( \frac{2}{3} - \frac{15}{4} x \ln \lb( \frac{2}{3x}\rb) + \frac{7}{2} x \rb).
\end{align}
Neglecting terms $\sim x^3$ (including logarithms) one has
\begin{align}
h_1 = \frac{\lambda x^2}{3}, && h_2 = - h_3 = \frac{41 \lambda x^2}{36} \lb( 3\ln \lb( \frac{2}{3x} \rb) -2 \rb).
\end{align}
For $\Delta = 0$ the consistency condition \eqref{S7B} is not met since
\begin{equation}
h_1 - \frac{2}{3} h_3 = \frac{\lambda x^2}{9} \lb( \frac{41}{2} \ln \lb( \frac{2}{3x} \rb) - \frac{32}{3}\rb)
\end{equation}
is of the order $x^2$, while $h_3^2$ is of the order $x^4$. We therefore need to include the effects of $\Delta \neq 0$.

We can employ the consistency condition for a determination of $\Delta$. In leading order $\Delta$ will be found to be of the order 
$x^2$. It is therefore sufficient to include in $h_1$ and $h_3$ the leading order terms
\begin{align}\label{S23}
\delta h_1 = - \frac{\lambda}{3} \Delta, \quad \delta h_2 = - \delta h_3 - \lambda \Delta.
\end{align}
The consistency condition \eqref{S7B} fixes $\Delta$ according to 
\begin{equation}
h_3^2 = 3\lambda \lb[ \frac{\lambda x^2}{9} \lb( \frac{41}{2}\ln \lb( \frac{2}{3x} \rb) -\frac{32}{3} \rb) + \frac{\lambda \Delta}{3} \rb].
\end{equation}
One infers
\begin{equation}\label{S25}
\Delta = - \lb( \frac{41}{6} \ln \lb(\frac{2}{3x} \rb) - \frac{32}{9} \rb) x^2 + \mathcal{O}(x^3).
\end{equation}
We find that $\Delta(t)$ is indeed a function of $\chi(t)$ such that the approximations \eqref{S3}, \eqref{S17A} are valid. Furthermore, with $h_3^2 \sim x^4$ only the contribution of $\delta h_k$ proportional to $x^2$ needs to be included in the leading terms for $h_k$. Adding the terms \eqref{S23} one finds the leading contributions
\begin{equation}
h_1 = \frac{\lambda x^2}{9} \left(\frac{41}{2}\ln \left(\frac{2}{3x}\right) -\frac{23}{3}\right),
\end{equation}
\begin{align}
h_3 = - h_2 = \frac{\lambda x^2}{6} \lb( \frac{41}{2} \ln \lb( \frac{2}{3x} \rb) - \frac{23}{3}\rb).
\end{align}
Both $h_1$ and $h_3$ are positive for small $x$.

For a check of the consistency condition \eqref{S25} for $\dot{H}$ we observe for $\Delta$ approximated by the leading order term \eqref{S25}
\begin{align}\label{AD1}
&\frac{1}{3} \frac{\del h_3}{\del \ln \chi} - \frac{h_3}{6} \frac{\del \ln \lambda}{\del \ln \chi}  
= \lambda x^3  \left(\frac{41}{9} \ln\left(\frac{2}{3x}\right) -\frac{215}{54}\right)  \nonumber \\
&- \quad \frac{\lambda x^4}{12} \left(41\ln \left(\frac{2}{3x}\right)-\frac{46}{3}\right) 
\end{align}
and
\begin{align}\label{AD2}
&h_2 + h_3 = -\frac{B\lambda}{6} (1 + \Delta) - \frac{1}{6} \frac{\del}{\del \ln \chi} [\lambda(1 + \Delta)] \nonumber \\
&= \lambda x^3 (1 + \Delta) \left(\frac{2}{3} - \frac{5x}{4}\ln \left(\frac{2}{3x}\right) + \frac{9x}{4}\right) -\frac{\lambda}{6} \frac{\del \Delta}{\del \ln \chi} \nonumber \\
&= \lambda x^3 \left[\frac{41}{9} \ln\left(\frac{2}{3x}\right) - \frac{215}{54} - \frac{5x}{4}\ln\left(\frac{2}{3x}\right) + \frac{9x}{4}\right].
\end{align}
The terms $\sim x^3$ are indeed identical for eqs.~\eqref{AD1} and \eqref{AD2}. For the terms $\sim x^4$ one would have to include corrections $\sim x^3$ in $\Delta$.

We can now determine the Hubble parameter from eq.~\eqref{S7A}
\begin{equation}\label{S27}
\frac{H}{\chi} = \frac{h_3}{\sqrt{3\lambda}} = \frac{\sqrt{\lambda}}{6\sqrt{3}}x^2 \lb( \frac{41}{2} \ln \lb( \frac{2}{3x} \rb) - \frac{23}{3} \rb). 
\end{equation}
For $t \to -\infty$ the Hubble parameter decreases to zero faster than $\chi$ since $x$ goes to zero. Insertion of the leading expression \eqref{S14} for $\chi(t)$, neglecting $\delta$ and keeping only the dominant logarithmic term in eq.~\eqref{S27} one finds
\begin{equation}
H = \frac{4\alpha(t)}{(t_0 - t)\ln^2\lb(\frac{\mu^2\lambda_0 (t_0 - t)^2}{3}\rb)},
\end{equation}
with a very slowly varying function
\begin{equation}\label{S29}
\alpha(t) = \frac{41}{48} \ln \lb[ \frac{2}{3} \ln \lb( \frac{\mu^2\lambda_0 (t_0 - t)^2}{3} \rb)\rb].
\end{equation}

For an approximate solution for the scale factor $a(t)$ we can neglect time derivatives of $\alpha(t)$. The solution of $\del_t \ln a = H$ then reads
\begin{equation}
a(t) = \bar{a} \exp \Bigg\{ \frac{\alpha(t)}{ \ln \lb( \sqrt{ \frac{\lambda_0}{3}} \mu (t_0 - t)\rb)} \Bigg\} .
\end{equation}
In the infinite past for $t\to -\infty$ the scale factor approaches the constant $\bar{a}$ with an inverse logarithm
\begin{equation}
a(t) = \bar{a} \Bigg( 1 + \frac{\alpha(t)}{\ln \lb(\sqrt{ \frac{\lambda_0}{3}} \mu (t_0 - t) \rb)} \Bigg).
\end{equation}
The geometry becomes flat Minkowski space. Eq.~\eqref{S29} specifies the function $\alpha(t)$ in eq.~\eqref{AF2}.

\section{General models with flat geometry in the infinite past}\label{Appendix B}

In this appendix we discuss a family of models for which the beginning epoch of the Universe is flat Minkowski space. They are based on an effective action for the metric and a scalar field containing no more than two derivatives. The Planck mass is given by the value of the scalar field $\chi$. The kinetic term for $\chi$ has a negative coefficient, without introducing any ghost or tachyon instability. These models are analogous in some respects to models of ``genesis" in higher derivative theories \cite{CREM1,CREM2,RUB1,RUB2}. We include in the discussion also scenarios where the curvature tensor vanishes in the infinite past, while the scale factor approaches zero instead of a non-zero constant.

For this purpose we consider the effective action \eqref{S1} for variable gravity and discuss the solutions of the field equations \eqref{eq:FR9}-\eqref{eq:FR8} or \eqref{F2} - \eqref{F4}. Different models are characterized by different functions $\lambda(\chi)$ and $B(\chi)$. We employ for small $\chi^2$ the shorthand
\begin{equation}\label{F5}
x = \frac{1}{\ln \left( \frac{\mu^2}{\chi^2}\right)},
\end{equation}
and assume a class of models which admit for small $\chi$ or small $x$ expansions of the type 
\begin{align}\label{F6}
\lambda = \frac{a_2}{x^2} + \frac{a_1}{x} + \lambda_0 + d_1 x + d_2 x^2 + d_3 x^3 + d_4 x^4, \nonumber \\
B = b_0 + b_1 x + b_2 x^2 + b_3 x^3 + b_4 x^4 + b_5 x^5.
\end{align}
For $\lambda = \lambda_0$, $B=b_0$ exact scale symmetry is realized. The other terms in the expansion therefore indicate derivations from scale symmetry. With
\begin{equation}\label{F7}
\frac{\del x}{\del \ln \chi} = 2 x^2,
\end{equation}
one has
\begin{align}
\frac{\del \lambda}{\del \ln \chi} &= - \frac{4 a_2}{x} - 2a_1 + 2 d_1 x^2 + 4 d_2 x^3 + 6 d_3 x^4 + 8 d_4 x^5, \nonumber \\
\frac{\del \ln B}{\del \ln \chi} &= \frac{2 b_1 x^2 + 4 b_2 x^3 + 6 b_3 x^4 + 8 b_4 x^5 + 10 b_5 x^6}{b_0 + b_1 x + b_2 x^2 + b_3 x^3 + b_4 x^4 + b_5 x^5}.
\end{align}

We want to investigate for which values of the coefficients in the expansion \eqref{F6} one can obtain Minkowski space as a solution of the field equations. This requires that the expressions $f_k$ in eqs.~\eqref{F2}-\eqref{F4} vanish fast enough for $x \to 0$. We first perform a systematic expansion in $x$. This computation will be rather tedious, since control over rather large powers of $x$ is needed. The result will later be compared to the primordial flat frame condition which gives a much easier access.

We obtain small values of $f_1$ if the two terms in eq.~\eqref{F2} cancel
\begin{equation}
\frac{\dot \chi^2}{\chi^4} \approx \frac{\lambda}{3(1 - \frac{B}{6})}.
\end{equation}
For the solution of the field equations we therefore make the ansatz
\begin{equation}\label{F10}
\frac{\dot \chi^2}{\chi^4} = c^2 (1 + e_1 x + e_2 x^2 + e_3 x^3) = \tilde{c}^2,
\end{equation}
with 
\begin{equation}
c^2 = \frac{\bar{\lambda}}{3(1-\frac{b_0}{6})},
\end{equation}
and $\bar{\lambda}$ the leading term in the expansion \eqref{F6}  for $\lambda$. For this ansatz we can employ
\begin{equation}
\ddot \chi = \dot \chi \frac{\del}{\del \chi} (\tilde{c}\chi^2),
\end{equation}
or
\begin{equation}
\frac{\ddot \chi}{\chi} = \frac{\dot \chi}{\chi^2} \frac{\del}{\del \ln \chi}(\tilde{c}\chi^2) = \tilde{c}^2 \left(2+ \frac{\del \ln \tilde{c}}{\del \ln \chi}\right)\chi^2,
\end{equation}
in order to bring $f_3$ into a form similar to $f_1$ and $f_2$. One obtains
\begin{equation}\label{F13A}
f_2 + f_3 = - \frac{1}{2} \left( B + \frac{\del \ln \tilde{c}^2}{\del \ln \chi} \right) \tilde{c}^2 \chi^2.
\end{equation}

Next we insert the ansatz \eqref{F10} in the coefficients 
\begin{equation}
f_k = \chi^2 h_k,
\end{equation}
with
\begin{align}
h_1 &= \frac{1}{3} \left[ \frac{a_2}{x^2} + \frac{a_1}{x} + \lambda_0 + d_1 x + d_2 x^2 + d_3 x^3 + d_4 x^4 \right. \nonumber\\
& - \bar{\lambda} \left(1 - \frac{1}{6} \frac{b_1 x + b_2 x^2 + b_3 x^3 + b_4 x^4 + b_5 x^5}{1- \frac{b_0}{6}}\right) \nonumber \\
& \times \left(1 + e_1 x + e_2 x^2 + e_3 x^3\right)\Bigr].
\end{align}
The leading term in $\lambda$ is canceled by $\bar{\lambda}$. For $h_2$ one has 
\begin{align}
h_2 &= \frac{-4a_2 x^{-1} - 2a_1 + 2d_1 x^2 +4 d_2 x^3 + 6 d_3 x^4 + 8 d_4 x^5}{b_0 + b_1 x + b_2 x^2 + b_3 x^3 + b_4 x^4 + b_5 x^5}\nonumber\\
& + \bar{\lambda}(1 + \mathcal{O}(x)).
\end{align}
We require that the highest power $\sim \bar{\lambda}$ is canceled as well. This requires
\begin{equation}\label{F16}
b_0 = 0,
\end{equation}
and further fixes $b_1$ as
\begin{align}\label{F17}
\bar{\lambda} &= \frac{a_2}{x^2}: && b_1 = 4, \nonumber \\
\bar{\lambda} &= \frac{a_1}{x}: && b_1 = 2, \nonumber \\
\bar{\lambda} &= \lambda_0: && b_1 = 0,\ b_2 = - \frac{2 d_1}{\lambda_0},  \nonumber \\
\bar{\lambda} &= d_1 x: && b_1 = -2.
\end{align}

We conclude that for modes of this type asymptotic flat space solutions exist only if $B$ vanishes for $\chi \to 0$, and $K = B-6$ is therefore negative. The leading coefficient of $B(x)$ is fixed by the leading power of $\lambda(x)$. For $h_3$ we employ
\begin{equation}\label{F18}
\frac{\ddot \chi}{\chi} = \left(2+ \frac{1}{2} \frac{\del \ln \tilde{c}^2}{\del \ln \chi}\right) \frac{\dot \chi^2}{\chi^2}
\end{equation}
and the observation that the leading term in $\del \ln(\tilde{c}^2)/\del\ln(\chi)$ is at most proportional to $x$. The leading order term in $h_3$ is therefore given by the negative leading order term of $h_2$, and no new constraint arises from the vanishing of the leading term in $h_3$.

The self-consistency of the three equations \eqref{F2} - \eqref{F4} can be expressed as a relation between the functions $h_k(x)$. The time derivative $\dot H$ can be computed both from a linear combination of eqs. \eqref{F3},\eqref{F4} and from the time derivative of $H^2$, which is a linear combination of eqs. \eqref{F2} and \eqref{F4}. This results in the relation
\begin{align}\label{F19A}
\del_t H^2 &= 2H\dot H = \del_t \left(f_1 - \frac{ 2 f_3}{3}\right) \nonumber\\
&= 2\left(h_1 - \frac{2 h_3}{3}\right)\chi \dot \chi + \del_t \left(h_1 - \frac{2 h_3}{3}\right)\chi^2 \nonumber\\
&= \frac{2}{3} \frac{f_3 \chi}{\dot\chi} \dot H. 
\end{align}
We can therefore express $\dot H$ as 
\begin{equation}
\dot H = \frac{3 \tilde{c}^2}{h_3} \left[h_1 - \frac{2 h_3}{3} + x^2 \frac{\del}{\del x}\left(h_1 - \frac{2 h_3}{3}\right)\right] \chi^2,
\end{equation}
and infer the consistency relation
\begin{align}\label{F19C}
&\chi^{-2} \left(\dot H - H \frac{\dot \chi}{\chi}\right) = h_2 + h_3 \nonumber \\
&= \frac{3 \tilde{c}^2}{h_3}\left[h_1 - \frac{2 h_3}{3} + x^2 \frac{\del}{\del x}\left(h_1 - \frac{2 h_3}{3}\right)\right]- \frac{h_3}{3}.
\end{align}

Let us concentrate on $\bar{\lambda} = \lambda_0$. Up to the order $x^3$ one has, with $\tilde{d}_k = d_k / \lambda_0$,
\begin{align}
h_1 &= \frac{\lambda_0}{3}\left[(\tilde{d}_1 - e_1)x + \left(\tilde{d}_2 - \frac{\tilde{d}_1}{3}-e_2\right)x^2\right. \nonumber\\
& + \left. \left(\tilde{d}_3 + \frac{b_2}{6} - \frac{\tilde{d}_1 e_1}{3} - e_3\right)x^3\right].
\end{align}
A similar expression for $h_2$ yields
\begin{align}
&h_2 = \lambda_0 \left[ \left( \frac{2}{3} + e_1 - \frac{2 \tilde{d}_2}{\tilde{d}_1} - \frac{b_3}{2 \tilde{d}_1}\right) x \right. \nonumber \\
&+ \left( e_2 + \frac{2}{3} e_1 + \frac{\tilde{d}_1}{3} - \frac{b_3}{6 \tilde{d}_1} - \frac{3 \tilde{d}_3}{\tilde{d}_1} - \frac{b_4}{2 \tilde{d}_1} - \frac{b_3^2}{4\tilde{d}_1^2} - \frac{\tilde{d}_2 b_3}{\tilde{d}_1^2}\right) x^2 \nonumber \\
& + c_3^{(2)}x^3\Biggr],
\end{align}
where
\begin{align}
c_3^{(2)} &= e_3 + \frac{2}{3} e_2 +\frac{e_1\tilde{d}_1}{3} - \frac{ b_3}{6} - \frac{b_4}{3\tilde{d}_1} - \frac{b_5}{2 \tilde{d}_1} - \frac{4\tilde{d}_4}{\tilde{d}_1} - \frac{e_1 b_3}{6\tilde{d}_1} \nonumber \\
& - \frac{b_3^2}{12 \tilde{d}_1} - \frac{3 b_3 \tilde{d}_3}{2 \tilde{d}_1^2} - \frac{b_4 \tilde{d}_2}{\tilde{d}_1^2} - \frac{b_3 b_4}{2 \tilde{d}_1^2} - \frac{b_3 \tilde{d}_2}{2\tilde{d}_1^3} - \frac{b_3^3}{8 \tilde{d}_1^3}
\end{align}
Here we use the expression
\begin{equation}
\frac{\del\ln B}{\del \ln \chi} = 4x - \frac{b_3}{\tilde{d}_1}x^2 - - \left(\frac{2 b_4}{\tilde{d}_1} + \frac{b_3^2}{2 \tilde{d}_1^2}\right)x^3. 
\end{equation}
For the remaining coefficient $h_3$ we use the relation \eqref{F13A}, which results in
\begin{equation}
h_2 + h_3 = -\frac{1}{2} \left( B + \frac{\del \ln \tilde{c}^2}{\del \ln\chi}\right)\tilde{c}^2.
\end{equation}
For $\bar{\lambda} = \lambda_0$ one obtains the expression in order $x^3$.
\begin{equation}
h_2 + h_3 = - \frac{\lambda_0}{3} \left[(e_1 - \tilde{d}_1)x^2 + \left(2 e_2 + \frac{b_3}{2} - e_1 \tilde{d}_1\right)x^3\right].
\end{equation}

For models that admit an expansion \eqref{F6} with $a_2 = a_1 = b_0 = b_1 = 0$ the ansatz \eqref{F10} for the solution implies that the r.h.s. of the field equations \eqref{F2}-\eqref{F4} vanishes at least $\sim \chi^2 x$ for $\chi \to 0$. Thus the Hubble parameter indeed vanishes in this limit. The asymptotic behavior of the scalar field for $t\to -\infty$ is given by the leading order term in the expression \eqref{F10}, $\dot \chi^2 = \lambda_0 \chi^4/3$, or
\begin{equation}\label{F25}
\frac{1}{\chi} = \sqrt{ \frac{\lambda_0}{3}} (t_0 - t) + \frac{1}{\chi_0}.
\end{equation}
Thus $\chi$ indeed vanishes $\sim (t_0 - t)^{-1}$, with 
\begin{equation}
\frac{\dot \chi}{\chi} = \sqrt{\frac{\lambda_0}{3}} \chi = \frac{1}{t_0-t}.
\end{equation}
For the last identity we have absorbed $\chi_0^{-1}$ in a shift of $t_0$. The divergence of $\chi$ for $t\to t_0$ is outside the range of validity of the approximation \eqref{F25}.

For the Hubble parameter eq.~\eqref{F4} implies 
\begin{equation}\label{F27}
H= \frac{h_3}{3\tilde{c}}\chi = \frac{h_3}{\lambda_0 (t_0-t)},
\end{equation}
while from eq.~\eqref{F2} one infers
\begin{equation}\label{F28}
H^2 = f_1 - \frac{2 f_3}{3} = \left(h_1-\frac{2h_3}{3}\right)\chi^2
\end{equation}
The consistency of the two equations \eqref{F27}\eqref{F28} requires
\begin{equation}
h_1 - \frac{2h_3}{3} = \frac{h_3^2}{9\tilde{c}^2}.
\end{equation}
For the leading order contribution one therefore has to require
\begin{equation}\label{F30}
\frac{h_1}{\lambda_0}-\frac{2h_3}{3\lambda_0} = \frac{h_3^2}{3 \lambda_0^2}.
\end{equation}
For the terms linear in $x$ in $h_1$ and $h_3$ one needs $3 h_1 = 2 h_3$. This fixes the coefficient $e_1$ for the solution as
\begin{equation}\label{F31}
e_1 = - \frac{4}{3} - \tilde{d}_1 + \frac{4 \tilde{d}_2}{\tilde{d}_1} + \frac{b_3}{\tilde{d}_1}.
\end{equation}
In turn, eq.~\eqref{F30} guarantees the consistency equation \eqref{F19C} in leading order. 

The detailed behavior of geometry for $t\to -\infty$ depends on the power of $x$ with which $h_3$ vanishes.
Let us first discuss the case where for $h_3$ the term linear in $x$ is different from zero
\begin{equation}
\frac{h_3}{\lambda_0} = 2 g_3 x.
\end{equation}
In this case one has
\begin{align}
H &= \frac{2g_3 x}{t_0-t} = \frac{g_3}{(t_0-t) \ln(\mu/\chi)}
\nonumber\\
&= \frac{g_3}{(t_0 - t) \ln \left(\sqrt{ \frac{\lambda_0}{3}} \mu(t_0 - t)\right)}.
\end{align}
The Hubble parameter, and therefore the curvature scalar, vanish for $t\to - \infty$. The corresponding evolution of the scale factor $a(t)$,
\begin{equation}
a(t) = \bar{a} \ln \left(\sqrt{ \frac{\lambda_0}{3} \mu(t_0 - t)}\right)^{-g_3}
\end{equation}
shows an inversely logarithmically vanishing scale factor $a(t\to -\infty)$.

For a setting where the scale factor approaches a constant value for $t\to -\infty$ we consider the leading behavior
\begin{equation}\label{F35}
\frac{h_3}{\lambda_0} = 4 \tilde{g}_3 x^2.
\end{equation}
With
\begin{equation}\label{F36}
H = \frac{4 \tilde{g}_3 x^2}{t_0 - t} = \frac{\tilde{g}_3}{(t_0-t)\ln^2 \left(\sqrt{ \frac{\lambda_0}{3}} \mu (t_0-t)\right)},
\end{equation}
the scale factor approaches a constant $\bar{a}$ inversely logarithmically for $t\to -\infty$
\begin{align}\label{F37}
a &= \bar{a} \exp \Bigg\{\frac{\tilde{g}_3}{\ln \left(\sqrt{ \frac{\lambda_0}{3}}\mu(t_0-t)\right) }\Bigg\}\nonumber\\
&\to \bar{a} \Bigg(1+ \frac{\tilde{g}_3}{\ln \left(\sqrt{ \frac{\lambda_0}{3}}\mu(t_0-t)\right)}\Bigg).
\end{align}
Geometry approaches Minkowski space with a constant scale factor $\bar{a}$ for $t\to -\infty$. We can identify $\alpha$ in eq.~\eqref{AF2} with the constant $\tilde{g}_3$.

For achieving eq.~\eqref{F35}, the term $\sim x$ in $h_3$ has to vanish. With $h_2 + h_3$ being at most of the order $x^2$ the condition is met if the term $\sim x$ vanishes for $h_2$. With eq.~\eqref{F31} we therefore require
\begin{align}\label{F38}
&\frac{2}{3} + e_1 - \frac{2 \tilde{d}_2}{\tilde{d}_1} - \frac{b_3}{2\tilde{d}_1} = \nonumber\\
& - \frac{2}{3} - \tilde{d}_1 + \frac{2\tilde{d}_2}{\tilde{d}_1} + \frac{b_3}{2 \tilde{d}_1} = 0.
\end{align}
This determines the coefficient $b_3$ in terms of $\tilde{d}_1$ and $\tilde{d}_2$, 
\begin{equation}
b_3 = \frac{4 \tilde{d}_1^2}{3} + 2\tilde{d}_1^2 - 4\tilde{d}_2.
\end{equation}
It also implies for the solution \eqref{F10}
\begin{equation}\label{F39}
e_1 = \tilde{d}_1. 
\end{equation}

Comparing eq.~\eqref{F28} with the square of eq.~\eqref{F36} yields the constraint
\begin{equation}\label{F40}
h_1 - \frac{2 h_3}{3} = \frac{16}{3} \tilde{g}_3^2 \lambda_0 x^4.
\end{equation}
Thus for the combination $h_1 - 2 h_3/3$ the terms $\sim x$, $\sim x^2$ and $\sim x^3$ all have to vanish. The expansion reads
\begin{align}
&h_1 - \frac{2h_3}{3} = h_1 + \frac{2h_2}{3} - \frac{2 (h_2 + h_3)}{3} \nonumber \\
&= \frac{\lambda_0}{3} \left[\left(\frac{4}{3} + e_1 + \tilde{d}_1 - \frac{4 \tilde{d}_2}{\tilde{d}_1} - \frac{b_3}{\tilde{d}_1}\right)x + s^{(2)}x^2 + s^{(3)}x^3\right],
\end{align}
with 
\begin{equation}
s^{(2)} = 2e_1 - e_2 - \frac{\tilde{d}_1}{3} + \tilde{d}_2 - \frac{b_3}{3\tilde{d}_1} - \frac{b_4}{\tilde{d}_1} - \frac{6\tilde{d}_3}{\tilde{d}_1}-\frac{2b_3\tilde{d}_2}{\tilde{d}_1^2}-\frac{b_3^2}{2 \tilde{d}_1^2},
\end{equation}
and
\begin{align}
s^{(3)} &= \frac{8}{3} e_2 + 2 e_3 - \frac{2b_4}{3\tilde{d}_1} - \frac{b_5}{\tilde{d}_1} - \frac{2 \tilde{d}_4}{\tilde{d}_1} - \frac{b_3 e_1}{3 \tilde{d}_1} - \frac{b_3^2}{6 \tilde{d}_1} \nonumber \\
&- \frac{3b_3 \tilde{d}_3}{\tilde{d}_1^2} - \frac{2 b_4 \tilde{d}_2}{\tilde{d}_1^2} - \frac{b_3 b_4}{\tilde{d}_1^2} - \frac{b_3 \tilde{d}_2}{\tilde{d}_1^3} - \frac{b_3^3}{4 \tilde{d}_1^3}.
\end{align}
The terms $\sim x$ vanishes by virtue of eq.~\eqref{F31}. Inserting eq.~\eqref{F39} into the constraint $s^{(2)} = 0$ determines the coefficient $e_2$ for the solution \eqref{F10},
\begin{equation}\label{F44}
e_2 = \frac{5\tilde{d}_1}{3} + \tilde{d}_2 - \frac{b_3}{3 \tilde{d}_1} - \frac{b_4}{\tilde{d}_1} - \frac{6 \tilde{d}_3}{\tilde{d}_1} - \frac{2b_3 \tilde{d}_2}{\tilde{d}_1^2} - \frac{b_3^2}{2 \tilde{d}_1^2}.
\end{equation}
Similarly, the coefficient $e_3$ is determined by $s^{(3)} = 0$. With $h_1 - 2h_3/3$ of the order $x^4$ and $h_3$ of the order $x^2$ the consistency condition \eqref{F19C} reads in the order $x^2$
\begin{equation}\label{F45}
h_2 + \frac{4 h_3}{3} = \frac{1}{\tilde{g}_3 x^2} \left(h_1 - \frac{2}{3}h_3\right).
\end{equation}
The r.h.s. of eq.~\eqref{F45} involves $e_4$.

For suitable coefficients $e_j$ in the expansion of the solution \eqref{F10} all consistency requirements are met. For $t\to -\infty$ the solution of the field equations \eqref{F2}-\eqref{F4} is indeed given by eq.~\eqref{F10}, with leading behavior \eqref{F25},\eqref{F36}, and geometry approaching flat Minkowski space according to eq.~\eqref{F37}. The coefficient $\tilde{g}_3$ is found as
\begin{equation}
\tilde{g}_3 = \frac{1}{8} \left(\tilde{d}_2 - \frac{1}{3}\tilde{d}_1 - 3e_2 \right).
\end{equation}

For the particular case
\begin{equation}\label{F47}
e_2 = \frac{1}{3} \tilde{d}_2 - \frac{1}{9} \tilde{d}_1
\end{equation}
the coefficient $\tilde{g}_3$ vanishes. The Hubble parameter approaches zero for $t\to -\infty$ even faster than for eq.~\eqref{F36}. Combining eqs.~\eqref{F44} and \eqref{F47} this is realized if the coefficients $b_k$ and $\tilde{d}_k$ obey an additional constraint. More generally, the closer the geometry is approximated by Minkowski space for large negative $t$, the more conditions on the coefficients of the expansions \eqref{F6} are needed.

The existence of the solutions which approach flat Minkowski space in the infinite past requires various conditions, as given bz eqs.~\eqref{F16},\eqref{F17}, and \eqref{F31} for $\bar{\lambda} = \lambda_0$. We may compare them with the condition for a primordial flat frame \eqref{eq:FR7}. With our ansatz \eqref{F6} this condition reads
\begin{equation}\label{F48}
B=2x^2 \left(\frac{1}{B-6} \frac{\del B}{\del x} - \frac{\del \ln \lambda}{\del x}\right).
\end{equation}
For $B-6$ different from zero the term $\sim \del B/\del x$ is subleading for $x\to 0$, such that in leading order in $x$ one has to require
\begin{equation}
\frac{\del \ln \bar{\lambda}}{\del \ln \chi} = - \frac{B}{2x}.
\end{equation}
For $x\to 0$ the l.h.s. is a constant or smaller. This requires $b_0 = 0$ and the relations \eqref{F18} for $b_1$. Expanding eq.~\eqref{F48} in powers of $x$ will lead to the constraints \eqref{F38}.

Indeed, for $\bar{\lambda} = \lambda_0$ the first terms of an expansion of eq.~\eqref{F48} in $x$ yields with $b_2 = -2 \tilde{d}_1$the relation
\begin{align}
B & -2x^2\left(\frac{1}{B-6} \frac{\del B}{\del x} - \frac{\del \ln \lambda}{\del x}\right) \nonumber \\
= & \left(b_3 - \frac{4}{3}\tilde{d}_1 + 4 \tilde{d}_2 -2\tilde{d}_1^2 \right)x^3 \nonumber\\
&+(b_4 + b_3 + 6\tilde{d}_3 - 6\tilde{d}_1\tilde{d}_2 + 2 \tilde{d}_1^3)x^4.
\end{align}
Setting the term $\sim x^3$ to zero yields the condition \eqref{F38}. If the term $\sim x^4$ also vanishes, one will find the condition that realizes the vanishing coefficient $\tilde{g}_3$ according to eq.~\eqref{F47}. Implementing the relation \eqref{F48} in a certain order in $x$ is the most direct way of establishing the conditions that lead to a decay of the Hubble parameter to zero with a certain power of $x$.

The discussion for a different leading behavior $\bar{\lambda}(x)$ can be performed in complete analogy. Minkowski space is approached in the infinite past if $H$ or $h_3$ vanish at least $\sim x^2$, and therefore $h_1-2h_3/3$ vanishes at least $\sim x^4$. These requirements again impose conditions on the coefficients of the expansion \eqref{F6}. The most direct way of finding these constraints is given by the expansion of eq.~\eqref{F48}, which has to be adapted to the leading behavior $\bar{\lambda}(x)$.

In summary, we have found a family of models for which geometry approaches Minkowski space in the infinite past. The cosmological solutions do not show any singularity. The models of this family all have $B\to 0$ as $t\to -\infty$. The value $B=0$ is the particular value for which the combination of curvature scalar term and scalar kinetic is invariant under conformal transformations, extending scale symmetry. For $B=0$ the scalar ceases to be a propagating degree of freedom.

The particular choice of models which admit an expansion \eqref{F6} in inverse powers of $\ln (\mu^2/\chi^2)$ may seem somewhat artificial at first sight. It is motivated by the observation that this type of model is obtained by expressing standard inflationary models in a different metric frame. For example, Starobinsky inflation corresponds to $a_2 = a_1=b_0=b_1=0$ and 
\begin{align}
\lambda_0 = \frac{1}{8C}, && \tilde{d}_1 = -3, && b_2 = 6
\end{align}
in lowest order.
The higher order terms for the primordial flat frame for Starobinsky inflation contain logarithms and do not admit the expansion \eqref{F6}.
For chaotic inflation with a quadratic potential $V=b^2M^2\sigma^2$ one has $a_2 = b_0 = 0$ and in lowest order $a_1 = b, b_1 = 2$.

It seems likely that the particular logarithmic behavior \eqref{F5} is not crucial for models with primordial flat space. It will be sufficient that the condition \eqref{eq:FR7} is obeyed with sufficient accuracy for some small quantity $x(\chi)$ that vanishes for $\chi \to 0$. This replaces the condition \eqref{F48} by
\begin{equation}
C_{pf} = B+\frac{\del x}{\del \ln \chi}\left(\frac{1}{6-B} \frac{\del B}{\del x} + \frac{\del \ln \lambda}{\del x}\right)\to 0.
\end{equation}
If $C_{pf}$ vanishes fast enough for $x\to 0$ the geometry will become Minkowski space in the infinite past.

\section{Chaotic inflation in the primordial flat frame}\label{Appendix C}

In this appendix we discuss in more detail the solutions of the field equations in the primordial flat frame for chaotic inflation. We remain in the leading order approximation for the constraint equation for the primordial flat frame. In this approximation we discuss the next to leading order of the solution for the time-dependence of the scalar field, and the associated evolution of geometry. One finds that the Hubble parameter and the curvature tensor vanish in the infinite past, while the scale factor approaches zero very slowly. Such a situation is close to Minkowski space, but exact Minkowski space is not reached in the infinite past. For a realization of Minkowski space in the infinite past one has to modify the field transformation in order to include next to leading order effects for the solution of the primordial flat frame constraint.

Let us start with the effective action
\begin{align}\label{W1}
\Gamma = \int_x & \sqrt{g} \Bigg\{-\frac{\chi^2}{2}R +b \chi^4 \ln \left(\frac{\mu^2}{\chi^2} + 1 \right) + \nonumber \\
& \left(\left[\ln \left(\frac{\mu^2}{\chi^2} + 1\right)\left(1+\frac{\chi^2}{\mu^2}\right)^2\right]^{-1} \del^\mu\chi \del_{\mu}\chi-3\right) \Bigg\}.
\end{align}
It contains up to two derivatives and has no instabilities as tachyons or ghosts despite a negative coefficient of the kinetic term. This model can be taken as a self-consistent model for all values of $\chi$. In the region of small $\chi$ it yields the model \eqref{eq:FR20A} of sect.~\ref{Inflationary_models_in_the_primordial_flat_frame}. For large $\chi$ one has a quadratic potential and vanishing $B$
\begin{equation}
V = \lambda \chi^4 = b \mu^2 \chi^2, \quad B = \frac{2 \mu^2}{\chi^2}.
\end{equation}
The transition between the two regimes corresponds in the Einstein frame to the end of inflation. We will solve the field equations of this model in the vicinity of $\chi = 0$.

In the Einstein frame this model is equivalent to chaotic inflation with a purely quadratic potential with mass term
\begin{equation}
m^2 = b M^2.
\end{equation}
The metric and canonical scalar field in the Einstein frame are given by
\begin{align}\label{W3}
g_{E,\mu\nu} = \frac{\chi^2}{M^2} g_{\mu \nu}, && \sigma^2 = 2 M^2 \ln(\frac{\mu^2}{\chi^2}+1).
\end{align}
Insertion of the field transformations \eqref{W3} into the effective action \eqref{W1} yields the effective action in the Einstein frame
\begin{equation}
\Gamma = \int_x \sqrt{g_E} \left\{-\frac{M^2}{2} R_E + \frac{1}{2} \del^\mu\sigma\del_\mu \sigma + \frac{m^2}{2}\sigma^2\right\}.
\end{equation}
This is the action used for chaotic inflation.

Early stages of inflation correspond to large $\sigma/M$ and therefore to small $\chi/\mu$. Neglecting corrections $\sim \chi^2/\mu^2$ the field equations derived from the effective action \eqref{W1} can be brought to the form
\begin{align}
\ddot \chi + 3H\dot \chi &= b \chi^3 \ln \left(\frac{\mu^2}{\chi^2}\right)-\frac{\dot\chi^2}{\chi}\left(1+\frac{1}{\ln\left(\frac{\mu^2}{\chi^2}\right)}\right) \label{W5}\\
H^2 &+ 2H \frac{\dot \chi}{\chi} = f_1, \label{W6} \\
\dot H &- 4H\frac{\dot\chi}{\chi} = f_2,\label{W7}
\end{align}
with
\begin{align}
f_1 &= \frac{\dot\chi^2}{\chi^2}\left(\frac{1}{3\ln\left(\frac{\mu^2}{\chi^2}\right)}-1\right) + \frac{b}{3} \chi^2 \ln\left(\frac{\mu^2}{\chi^2}\right),
\\
f_2 &= 3 \frac{\dot\chi^2}{\chi^2} - b \chi^2 \ln\left(\frac{\mu^2}{\chi^2}\right).
\end{align}

The field equations have a simple asymptotic solution for $t\to -\infty$ where spacetime is flat, $H=0$, $\dot H = 0$, and $\chi$ approaches zero according to the implicit equation
\begin{equation}
\frac{1}{\chi} = c(\chi)(t_0-t) + \frac{1}{\chi_0},
\end{equation}
where $c$ increases logarithmically for $\chi \to 0$,
\begin{equation}
c^2 = \frac{b}{3}\left[\ln \left(\frac{\mu^2}{\chi^2}\right)+h\right].
\end{equation}
We will discuss the choice of the constant $h$ later. It is subleading for $\chi \to 0$. For this solution one finds in leading order in an expansion in $1/\ln(\mu/\chi)$ the relation
\begin{equation}\label{W13}
\left(\frac{\dot\chi}{\chi}\right)^2 = c^2 \chi^2 = \frac{b}{3}\chi^2 \ln\left(\frac{\mu^2}{\chi^2}\right),
\end{equation}
such that for $\chi\to 0$ both $f_1$ and $f_2$ vanish and eqs.~\eqref{W6},\eqref{W7} are obeyed for flat space.

As $\chi$ increases, geometry deviates from flat space. For a quantitative investigation we make the ansatz
\begin{equation}\label{W14}
\frac{1}{\chi} = c(t_0-t)+\delta,
\end{equation}
which implies
\begin{equation}
\frac{\dot\chi}{\chi} = \frac{(c-\dot\delta)\chi}{1+\frac{\del \ln c}{\del \ln \chi}(1-\delta\chi)}.
\end{equation}
Using
\begin{equation}
\frac{\del \ln c}{\del \ln \chi} = - \frac{1}{\ln \left(\frac{\mu^2}{\chi^2}\right)+h},
\end{equation}
we expand in powers of $1/\ln (\mu/\chi)$ and $\delta$, $\dot \delta$. This yields
\begin{equation}\label{W17}
\frac{\dot\chi}{\chi} = c \chi \left(1 + \frac{1}{\ln \left(\frac{\mu^2}{\chi^2}\right)}\right) -\dot \delta \chi,
\end{equation}
and
\begin{align}
f_1 &= - \frac{b}{6}\chi^2\left(\frac{10}{3}+ 2 h \right) + 2 c \chi^2 \dot \delta, \\
f_2 &= b \chi^2 (2+h) - 6 c \chi^2 \dot \delta.
\end{align}
The scalar field equation \eqref{W5} becomes
\begin{equation}\label{W18A}
3 H \frac{\dot\chi}{\chi} = f_3,
\end{equation}
where
\begin{equation}
f_3 = -b\chi^2(2+h) + 6c\chi^2 \dot\delta + \chi \ddot \delta.
\end{equation}

So far the quantities $f_1$, $f_2$, $f_3$ still involve the next to leading correction $\delta$ to the evolution of the scalar field. We therefore have to determine $\delta(t)$
For a solution of the field equation for $\delta$ we make the ansatz (with constant $A$)
\begin{equation}\label{W20}
\dot \delta = \frac{Ab}{c}.
\end{equation}
This implies that the term $\chi \ddot \delta$ is subleading and can be neglected as compared to the other terms in $f_3$,
\begin{equation}\label{W21}
\chi \ddot \delta = \frac{Ab\chi^2}{\ln\left(\frac{\mu^2}{\chi^2}\right)}.
\end{equation}
With eq.~\eqref{W20} one obtains
\begin{align}
f_1 &= -\frac{b}{6} \chi^2 \left(\frac{10}{3} + 2h - 12 A\right) = h_1 \chi^2, \nonumber \\
f_3 &= -f_2 = -b\chi^2 (2+h-6A) = h_3 \chi^2.
\end{align}
At this point the next to leading contribution $\delta$ to the scalar field enters only through the constant $A$.

We are now ready to extract the next to leading order for the metric.
From eqs.~\eqref{W18A},\eqref{W17}, one concludes
\begin{align}\label{C25}
H= \frac{h_3 \chi}{3c}, && \dot H = \frac{h_3 \dot \chi}{3c} \left( 1 - \frac{\del\ln c}{\del\ln\chi} \right),
\end{align}
which yields in lowest order
\begin{equation}\label{C26}
\dot H = H \frac{\dot \chi}{\chi}.
\end{equation}
Equations \eqref{W7} and \eqref{W18A}
are therefore identical.
Both $H$ and $\dot H$ go to zero for $\chi \to 0$, as expected for the primordial flat frame.

We still need to fix the constant $A$ in the next to leading behavior for the scalar.
For eq.~\eqref{W5} we observe that $H^2\sim \chi^2/c^2$ is subleading as compared to $H \dot \chi /\chi \sim \chi^2$. Thus the combination of eqs.~\eqref{W5} and \eqref{W18A} requires
\begin{equation}\label{C27}
f_1  = \frac{2 f_3}{3}, \quad h_1 = \frac{2 h_3}{3},
\end{equation}
or
\begin{equation}\label{W27}
6A-h = \frac{7}{3}, \quad h_3 = \frac{b}{3}.
\end{equation}
We observe that only the combination $6A-h$ is determined. Both enter the ansatz for the subleading correction to the evolution equation for $\chi$ in eq.~\eqref{W13}. Since we are only interested in the leading part of this correction we can choose $h$ freely as long as eq.~\eqref{W27} is obeyed. We take $h = -7/3$, such that in the next to leading order one has $A=0$ and therefore $\delta = 0$ in eq.~\eqref{W14} This establishes the implicit equation
\begin{equation}\label{C29}
\frac{1}{\chi^2} = \frac{b}{3}\left(\ln \left(\frac{\mu^2}{\chi^2}\right)-\frac{7}{3}\right)(t_0-t)^2.
\end{equation}
As compared with eq.~\eqref{eq:FR23} the next to leading correction replaces the parameter $\chi_0^{-1}$ by a fixed expression.
For a rough estimate of the logarithm we can employ the leading order for $\chi \to 0$ 
\begin{equation}
\ln \left(\frac{\mu^2}{\chi^2}\right) - \frac{7}{3} \approx \ln \left(\frac{b \mu^2(t_0 - t)^2}{3}\right).
\end{equation}
We arrive at the next to leading expression of the scalar field for $t\to \infty$,
\begin{equation}
\frac{1}{\chi^2} \approx \frac{b}{3} \ln \left(\frac{b\mu^2(t_0 - t)^2}{3}\right)(t_0-t)^2.
\end{equation}

For the Hubble parameter eqs.~\eqref{C25},\eqref{W27},\eqref{C29} imply
\begin{equation}
H^2 = \frac{b\chi^2}{27\left(\ln\left(\frac{\mu^2}{\chi^2}\right)-\frac{7}{3}\right)}.
\end{equation}
Due to the increase of $\ln(\mu^2/\chi^2)$ for $t\to -\infty$ the Hubble parameter decreases faster than $(t_0-t)^{-1}$ in this limit,
\begin{align}
H &= \frac{1}{\left(3\ln\left(\frac{\mu^2}{\chi^2}\right)-7\right)(t_0 - t)} \nonumber \\
&\approx \frac{1}{3\ln(b\mu^2 (t_0-t)^2/3)(t_0 - t)}.
\end{align}
For $t\to -\infty$ the scale factor approaches zero very slowly
\begin{equation}\label{C34}
a = \frac{\bar{a}}{[\ln(b\mu^2(t_0-t)^2/3)]^{1/6}}.
\end{equation}

For the effective action \eqref{W1}, based on the field transformation \eqref{W3}, the approach of the Hubble parameter to zero is not fast enough for achieving Minkowski space for $t \to -\infty$. While the evolution \eqref{C34} is very slow, the scale factor nevertheless decreases asymptotically to zero. Minkowski space remains, however, a rather good approximation for large time intervals. If one wants a formulation of the primordial flat frame for which Minkowski space is reached for $t\to -\infty$, one needs a solution of eq.~\eqref{CI2} beyond the leading order \eqref{eq:FR20} or \eqref{W3}. In the language of appendix \ref{Appendix B} the effective action \eqref{W1} corresponds to $\bar{\lambda} = b/x$, $B=2x$, $a_1 = b$, $b_1 = 2$. ``Improving" the primordial flat frame by continuing the expansion in higher order in $x$ will lead to solutions that approach Minkowski space in the infinite past.

\section{Primordial flat frame for Starobinsky \\ inflation}\label{Appendix D}

In this appendix we establish the equivalence of Starobinsky inflation with the variable gravity model discussed in sect.~\ref{Beginning_with flat_spacetime} and appendix \ref{Appendix A}. For this purpose we solve eq.~\eqref{SI4} in next to leading order. With the ansatz
\begin{equation}
B = 2\varepsilon (1+f_B)^2,
\end{equation}
one finds
\begin{equation}\label{D2}
(3 - \varepsilon(1 + f_B)^2)\sqrt{2\varepsilon} f_B = \frac{\del\varepsilon}{\del \tsigma}(1+f_B)^2 + 2\varepsilon(1+f_B)\frac{\del f_B}{\del\tsigma},
\end{equation}
with
\begin{equation}
\frac{\del\varepsilon}{\del\tsigma} = -\left(\sqrt{\frac{8}{3}} + \sqrt{2\varepsilon}\right)\varepsilon.
\end{equation}
In leading order $f_B$ and $\del f_B / \del \tsigma$ are proportional $\sqrt{\veps}$ and eq.~\eqref{D2} simplifies to
\begin{equation}
3 \sqrt{2\veps} f_B = \frac{\del \veps}{\del \tsigma} = -\sqrt{ \frac{8}{3}}\veps,
\end{equation}
or
\begin{equation}
f_B = - \frac{2}{3\sqrt{3}} \sqrt{\veps}.
\end{equation}

In this order the differential equation relating $\tsigma$ and $\chi$ reads
\begin{align}\label{60F}
\left(\frac{\del\tsigma}{\del\ln\chi}\right)^2 = B = 2\veps + 4\veps f_B = 2\veps - \frac{8}{3\sqrt{3}} \veps \sqrt{\veps}.
\end{align}
We introduce the shorthand
\begin{equation}
W = \exp \left(-\sqrt{\frac{2}{3}}\tsigma\right),
\end{equation}
with next to leading expression
\begin{align}
B &= \frac{8}{3}\left(\frac{W}{1-W}\right)^2 - \frac{64}{27} \left(\frac{W}{1-W}\right)^3 \nonumber \\
&= \frac{8}{3} W^2 + \frac{80}{27} W^3.
\end{align}
With 
\begin{equation}
\tilde{\gamma} = \sqrt{ \frac{2}{3}} \tsigma = - \ln{W},
\end{equation}
eq.~\eqref{60F} becomes
\begin{align}
\frac{\del \tilde{\gamma}}{\del \ln \chi} &= \sqrt{ \frac{2}{3}} \frac{\del \tsigma}{\del \ln \chi} = - \sqrt{ \frac{2}{3}} \sqrt{ \frac{8}{3} W^2 \left(1+\frac{10}{9}W\right) }\nonumber \\
&= - \frac{4}{3} e^{-\tilde{\gamma}}\left(1 + \frac{5}{9} e^{-\tilde{\gamma}}\right).
\end{align}

The next to leading order solution is given by
\begin{equation}
e^{\tilde{\gamma}}-\frac{5}{9} \tilde{\gamma} + c_0 = \frac{2}{3} \ln \left(\frac{\mu^2}{\chi^2}\right).
\end{equation}
This expresses $\tsigma$ or $W$ as a function of $\chi$ and one finds by an iterative solution for small $x$

\begin{align}\label{60L}
W &= e^{-\tilde{\gamma}} = \frac{3x}{2}\left(1 - \frac{5}{9}\tilde{\gamma} e^{-\tilde{\gamma}} + c_0 e^{-\tilde{\gamma}} \right)\nonumber\\
&= \frac{3x}{2}\left(1-\frac{5}{6}x \ln \left(\frac{2}{3x}\right)+ \frac{3 c_0 x}{2}\right),
\end{align}
with
\begin{align}
x = \frac{1}{\ln \left(\frac{\mu^2}{\chi^2}\right)}, && \frac{\del x}{\del\ln \chi} = 2 x^2.
\end{align}

For small $\chi$ and therefore small $x$ we can express $B$ in terms of $x$
\begin{equation}
B = 6 x^2 \left[1 - \frac{5}{3} x \left(\ln \left(\frac{2}{3x}\right) + \left(\frac{5}{3} + 3c_0\right)x\right)\right].
\end{equation}
Due to the logarithm in the relation \eqref{60L} between $W$ and $x$ a Taylor expansion of $B$ is possible in $W$, but not in $x$. Also $\lambda$ has a simple expression in terms of $W$,
\begin{align}
\lambda = \lambda_0 (1-W)^2, && \lambda_0 = \frac{1}{8C}.
\end{align}

We may check the primordial flat frame condition \eqref{eq:FR7} by computing
\begin{align}
C_{pf} &= B + \frac{1}{6-B} \frac{\del B}{\del \ln \chi} + \frac{\del \ln \lambda}{\del \ln \chi}\nonumber \\
&= B + \frac{\del W}{\del\ln\chi} \left(\frac{1}{6-B} \frac{\del B}{\del W} + \frac{\del \ln \lambda}{\del W}\right).
\end{align}
The combination $C_{pf}$ should vanish if the condition \eqref{eq:FR7} is obeyed.
We employ
\begin{align}
\frac{\del W}{\del \ln \chi} &= 2 x^2 \frac{\del W}{\del x} \nonumber \\
&= 3 x^2 \left(1 - \frac{5}{6}x \ln \left(\frac{2}{3x}\right) + \frac{3c_0 x}{2}\right)^2 \left(1+ \frac{5}{6}x\right) \nonumber \\
&= \frac{4}{3}W^2\left(1 + \frac{5}{9} W\right),
\end{align}
and
\begin{align}
\frac{1}{6-B} \frac{\del B}{\del W} &= \frac{8}{9} W \left(1 + \frac{5}{9}W\right), \nonumber\\
\frac{\del \ln \lambda}{\del W} &= - \frac{2}{1-W} = - 2(1+W + W^2),
\end{align}
in order to establish that $C_{pf}$ indeed vanishes up to terms $\sim W^4$. We observe that the integration constant $c_0$ only affects the relation between $W$ and $x$. It drops out in the relation between $W$ and $\chi$. Setting $c_0 = 0$ we obtain the effective action \eqref{S1},\eqref{S2} discussed in sect.~\ref{Beginning_with flat_spacetime}. This establishes that this effective action corresponds to a particular choice of metric, namely the primordial flat frame, for Starobinsky inflation.

\section{Primordial flat frame for scaling solutions in quantum gravity}\label{Appendix E}

In this appendix we derive the effective action in the primordial flat frame for the scaling solutions \eqref{87}-\eqref{89} for the functional flow in quantum gravity. With the Weyl scaling \eqref{91} one obtains
the effective action \eqref{S1} with
\begin{equation}
\lambda = \frac{\bar{U}}{\bar{F}^2} = \frac{u}{4w^2} = \frac{u_0 + \tilde{m}^2 \trho}{(2w_0 + \xi \trho)^2},
\end{equation}
and
\begin{align}
K = \left[\frac{\bar{K}k^2}{2\bar{F}\trho} + \frac{3}{2} \left(\frac{\del \ln \bar{F}}{\del\trho}\right)^2\right]\left(\frac{\del\trho}{\del\ln\chi}\right)^2 -6, 
\end{align}
or
\begin{align}
B = K+6 = \left[\frac{\kappa}{4\tilde{w}(\trho)\trho} + \frac{3}{2} \left(\frac{\del\ln \tilde{w}(\trho)}{\del\trho}\right)^2\right]\left(\frac{\del\trho}{\del\ln\chi}\right)^2.
\end{align}
If $\del \ln \bar{F}/\del \ln \trho$ can be neglected, one has
\begin{equation}
B = \frac{\bar{K}}{\bar{F}}\left(\frac{\del \psi}{\del\ln\chi}\right)^2,
\end{equation}
in accordance with eq.~\eqref{eq:FR4} for $\bar{K}=1$, $\bar{F} = M^2$, $\psi = \sigma$.
More generally, we write
\begin{align}
B &= \beta(\trho) \left(\frac{\del \trho}{\del\ln\chi}\right)^2, \nonumber\\
\beta(\trho) &= \frac{\kappa}{4\tilde{w}(\trho)\trho} + \frac{3}{2}\left(\frac{\del\ln \tilde{w}(\trho)}{\del \trho}\right)^2 \nonumber \\
&= \frac{\kappa_0 +\kappa_1 \trho}{2\trho (2w_0 + \xi \trho)} + \frac{3 \xi^2}{2(2w_0 + \xi \trho)^2}.
\end{align}
For the scaling solution the expressions of $\lambda$ and $B$ in terms of the dimensionless invariant $\trho$ are independent of $k$.
Away from the scaling solution $\lambda$ and $B$ will additionally depend on some intrinsic mass scale.

For a transformation to the primordial flat frame one needs to find the relation $\trho(\chi)$ such that the condition \eqref{eq:FR7} is satisfied
\begin{equation}
B = \frac{1}{B-6} \frac{\del B}{\del \ln \chi} - \frac{\del \ln \lambda}{\del\ln \chi}.
\end{equation}
This can be written as a differential equation for $B(\trho)$,
\begin{equation}
\sqrt{\beta B} = \frac{1}{B-6} \frac{\del B}{\del \trho} - \frac{\del \ln \lambda}{\del \trho},
\end{equation}
where we have assumed $\del \trho/\del \ln \chi > 0$. Let us take models with
\begin{equation}
- \frac{\del \ln \lambda}{\del \trho} = \frac{2\xi}{2w_0 + \xi \trho} - \frac{\tilde{m}^2}{u_0 + \tilde{m} \trho} >0.
\end{equation}
With 
\begin{align}
\beta &= \frac{\bar{\beta}}{\trho}, \quad \bar{\beta} = \frac{\kappa_0}{2(2w_0 + \xi \trho)} (1+c_1 \trho), \nonumber \\
c_1 &= \frac{\kappa_1}{\kappa_0} + \frac{3\xi^2}{\kappa_0(2w_0 + \xi \trho)},
\end{align}
we have in lowest order of $\trho$
\begin{equation}
B=b_1 \trho,
\end{equation}
where $b_1 >0$ is determined by
\begin{equation}\label{QT6}
\sqrt{ \frac{\kappa_0 b_1}{4 w_0}} = - \frac{b_1}{6} + \frac{\xi}{w_0} - \frac{\tilde{m}^2}{u_0}.
\end{equation}
The solution with positive $b_1$ reads
\begin{equation}\label{QT7}
b_1 = 6a + \frac{9 \kappa_0}{2 w_0}\left(1 \pm \sqrt{1 + \frac{8 a w_0}{3 \kappa_0}}\right),
\end{equation}
where
\begin{equation}
a = \frac{\xi}{w_0} - \frac{\tilde{m}^2}{u_0}.
\end{equation}
For $\kappa_0 / w_0 >0$ the r.h.s. of eq.~\eqref{QT6} is positive if the minus sign applies in eq.~\eqref{QT7}, and if $a$ is positive. Positive $a$ is therefore a necessary condition for the existence of the map to the primordial flat frame with positive $\del \trho / \del \ln \chi$. We also have to require $b_1 > 0$. This is indeed the case.

In the same leading approximation one has
\begin{equation}
\frac{\del \trho}{\del \ln \chi} = \sqrt{\frac{B}{\beta}} = \gamma \trho,
\end{equation} 
with
\begin{equation}
\gamma = \sqrt{ \frac{4 b_1 w_0}{\kappa_0}} = \left( \frac{24 w_0 a}{\kappa_0} + 18 \left(1 - \sqrt{1+ \frac{8 a w_0}{3 \kappa_0}}\right)\right)^{\frac{1}{2}}.
\end{equation}
This establishes the leading order relation between $\trho$ and $\chi$,
\begin{equation}\label{QT11}
\trho = \left(\frac{\chi}{\tilde{\mu}}\right)^\gamma, \quad \chi = \tilde{\mu} \trho^{\frac{1}{\gamma}},
\end{equation}
with $\tilde{\mu}$ an integration constant.
In turn, this yields the effective action \eqref{S1} in the primordial flat frame according to eqs.~ \eqref{106}-\eqref{108}.
For $8aw_0 \ll 3\kappa_0$ one has the approximative expressions
\begin{equation}
b_1 = \frac{4 a^2 w_0}{\kappa_0}, \quad \gamma = \frac{4 a w_0}{\kappa_0}.
\end{equation}

\nocite{*}
\bibliography{PFF_refs}

\end{document}